\documentclass[journal, onecolumn, 12pt]{IEEEtran}

\usepackage{graphics}
\usepackage{rotating}
\usepackage{amsfonts}
\usepackage{amsmath}
\usepackage{subfigure}
\usepackage{verbatim}
\usepackage{enumerate}
\usepackage{setspace}

\doublespace
\centerfigcaptionstrue

\usepackage{epsfig}
\graphicspath{{figures/}} \DeclareGraphicsExtensions{.eps,.ps}

\begin{document}

\title{Analysis of the 802.11e Enhanced Distributed Channel Access
Function \footnotemark{$^{\dag}$}}

\author{\singlespace \normalsize \authorblockN{Inanc Inan, Feyza Keceli, and Ender Ayanoglu}\\
\authorblockA{Center for Pervasive Communications and Computing \\
Department of Electrical Engineering and Computer Science\\
The Henry Samueli School of Engineering\\
University of California, Irvine, 92697-2625\\
Email: \{iinan, fkeceli, ayanoglu\}@uci.edu}}

\maketitle

\footnotetext{$^{\dag}$ This work is supported by the Center for
Pervasive Communications and Computing, and by National Science
Foundation under Grant No. 0434928. Any opinions, findings, and
conclusions or recommendations expressed in this material are
those of authors and do not necessarily reflect the view of the
National Science Foundation.}

\begin{abstract} The IEEE 802.11e standard revises the Medium
Access Control (MAC) layer of the former IEEE 802.11 standard for
Quality-of-Service (QoS) provision in the Wireless Local Area
Networks (WLANs). The Enhanced Distributed Channel Access (EDCA)
function of 802.11e defines multiple Access Categories (AC) with
AC-specific Contention Window (CW) sizes, Arbitration Interframe
Space (AIFS) values, and Transmit Opportunity (TXOP) limits to
support MAC-level QoS and prioritization. We propose an analytical
model for the EDCA function which incorporates an accurate CW,
AIFS, and TXOP differentiation at any traffic load. The proposed
model is also shown to capture the effect of MAC layer buffer size
on the performance. Analytical and simulation results are compared
to demonstrate the accuracy of the proposed approach for varying
traffic loads, EDCA parameters, and MAC layer buffer space.
\end{abstract}

\section{Introduction}\label{sec:intro}

The IEEE 802.11 standard \cite{802.11} defines the Distributed
Coordination Function (DCF) which provides best-effort service at
the Medium Access Control (MAC) layer of the Wireless Local Area
Networks (WLANs). The recently ratified IEEE 802.11e standard
\cite{802.11e} specifies the Hybrid Coordination Function (HCF)
which enables prioritized and parameterized Quality-of-Service
(QoS) services at the MAC layer, on top of DCF. The HCF combines a
distributed contention-based channel access mechanism, referred to
as Enhanced Distributed Channel Access (EDCA), and a centralized
polling-based channel access mechanism, referred to as HCF
Controlled Channel Access (HCCA).

We confine our analysis to the EDCA scheme, which uses Carrier
Sense Multiple Access with Collision Avoidance (CSMA/CA) and
slotted Binary Exponential Backoff (BEB) mechanism as the basic
access method. The EDCA defines multiple Access Categories (AC)
with AC-specific Contention Window (CW) sizes, Arbitration
Interframe Space (AIFS) values, and Transmit Opportunity (TXOP)
limits to support MAC-level QoS and prioritization \cite{802.11e}.

In order to assess the performance of these functions, simulations
or mathematical analysis can be used. Although simulation models
may capture system dynamics very closely, they lack explicit
mathematical relations between the network parameters and
performance measures. A number of networking functions would
benefit from the insights provided by such mathematical relations.
For example, analytical modeling is a more convenient way to
assist embedded QoS-aware MAC scheduling and Call Admission
Control (CAC) algorithms. Theoretical analysis can provide
invaluable insights for QoS provisioning in the WLAN. On the other
hand, analytical modeling can potentially be complex, where the
effect of multiple layer network parameters makes the task of
deriving a simple and accurate analytical model highly difficult.
However, a set of appropriate assumptions may lead to simple yet
accurate analytical models.

The majority of analytical work on the performance of 802.11e EDCA
(and of 802.11 DCF) assumes that every station has always
backlogged data ready to transmit in its buffer anytime (in
saturation) as will be discussed in Section~\ref{sec:related}.
Analysis of the system in this state (saturation analysis)
provides accurate and practical asymptotic figures. However, the
saturation assumption is unlikely to be valid in practice given
the fact that the demanded bandwidth for most of the Internet
traffic is variable with significant idle periods. Our main
contribution is an accurate EDCA analytical model which releases
the saturation assumption. The model is shown to predict EDCA
performance accurately for the whole traffic load range from a
lightly loaded non-saturated channel to a heavily congested
saturated medium for a range of traffic models.

Similarly, the majority of analytical work on the performance of
802.11e EDCA (and of 802.11 DCF) assumes constant collision
probability for any transmitted packet at an arbitrary backoff
slot independent of the number of retransmissions it has
experienced. A complementary assumption is the constant
transmission probability for any AC at an arbitrary backoff slot
independent of the number of retransmissions it has experienced.
As will be discussed in Section~\ref{sec:related}, these
approximations lead to accurate analysis in saturation. Our
analysis shows that the slot homogeneity assumption leads to
accurate performance prediction even when the saturation
assumption is released.

Furthermore, the majority of analytical work on the performance of
802.11e EDCA (and of 802.11 DCF) in non-saturated conditions
assumes either a very small or an infinitely large MAC layer
buffer space. Our analysis removes such assumptions by
incorporating the finite size MAC layer queue (interface queue
between Link Layer (LL) and MAC layer) into the model. The finite
size queue analysis shows the effect of MAC layer buffer space on
EDCA performance which we will show to be significant.

A key contribution of this work is that the proposed analytical
model incorporates \textit{all} EDCA QoS parameters, CW, AIFS, and
TXOP. The model also considers varying collision probabilities at
different AIFS slots which is a direct result of varying number of
contending stations. Comparing with simulations, we show that our
model can provide accurate results for an arbitrary selection of
AC-specific EDCA parameters at any load.


We present a Markov model the states of which represent the state
of the backoff process and MAC buffer occupancy. To enable
analysis in the Markov framework, we assume constant probability
of packet arrival per state (for the sake of simplicity, Poisson
arrivals). On the other hand, we have also shown that the results
also hold for a range of traffic types.

\section{EDCA Overview}\label{sec:EDCAoverview}

The IEEE 802.11e EDCA is a QoS extension of IEEE 802.11 DCF. The
major enhancement to support QoS is that EDCA differentiates
packets using different priorities and maps them to specific ACs
that are buffered in separate queues at a station. Each AC$_{i}$
within a station ($0\leq i\leq i_{max}$, $i_{max}=3$ in
\cite{802.11e}) having its own EDCA parameters contends for the
channel independently of the others. Following the convention of
\cite{802.11e}, the larger the index $i$ is, the higher the
priority of the AC is. Levels of services are provided through
different assignments of the AC specific EDCA parameters; AIFS,
CW, and TXOP limits.

If there is a packet ready for transmission in the MAC queue of an
AC, the EDCA function must sense the channel to be idle for a
complete AIFS before it can start the transmission. The AIFS of an
AC is determined by using the MAC Information Base (MIB)
parameters as
\begin{align}
AIFS = SIFS + AIFSN \times T_{slot},
\end{align}

\noindent where $AIFSN$ is the AC-specific AIFS number, $SIFS$ is
the length of the Short Interframe Space and $T_{slot}$ is the
duration of a time slot.

If the channel is idle when the first packet arrives at the AC
queue, the packet can be directly transmitted as soon as the
channel is sensed to be idle for AIFS. Otherwise, a backoff
procedure is completed following the completion of AIFS before the
transmission of this packet. A uniformly distributed random
integer, namely a backoff value, is selected from the range
$[0,W]$. Should the channel be sensed busy at any time slot during
AIFS or backoff, the backoff procedure is suspended at the current
backoff value. The backoff resumes as soon as the channel is
sensed to be idle for AIFS again. When the backoff counter reaches
zero, the packet is transmitted in the following slot.

The value of $W$ depends on the number of retransmissions the
current packet experienced. The initial value of $W$ is set to the
AC-specific $CW_{min}$. If the transmitter cannot receive an
Acknowledgment (ACK) packet from the receiver in a timeout
interval, the transmission is labeled as unsuccessful and the
packet is scheduled for retransmission. At each unsuccessful
transmission, the value of $W$ is doubled until the maximum
AC-specific $CW_{max}$ limit is reached. The value of $W$ is reset
to the AC-specific $CW_{min}$ if the transmission is successful,
or the retry limit is reached thus the packet is dropped.

The higher priority ACs are assigned smaller AIFSN. Therefore, the
higher priority ACs can either transmit or decrement their backoff
counters while lower priority ACs are still waiting in AIFS. This
results in higher priority ACs enjoying a lower average
probability of collision and relatively faster progress through
backoff slots. Moreover, in EDCA, the ACs with higher priority may
select backoff values from a comparably smaller CW range. This
approach prioritizes the access since a smaller CW value means a
smaller backoff delay before the transmission.

Upon gaining the access to the medium, each AC may carry out
multiple frame exchange sequences as long as the total access
duration does not go over a TXOP limit. Within a TXOP, the
transmissions are separated by SIFS. Multiple frame transmissions
in a TXOP can reduce the overhead due to contention. A TXOP limit
of zero corresponds to only one frame exchange per access.

An internal (virtual) collision within a station is handled by
granting the access to the AC with the highest priority. The ACs
with lower priority that suffer from a virtual collision run the
collision procedure as if an outside collision has occured
\cite{802.11e}.

\section{Related Work}\label{sec:related}

In this section, we provide a brief summary of the theoretical DCF
and EDCA function performance analysis in the literature.

The majority of previous work carries out performance analysis for
asymptotical conditions assuming each station is in saturation.
Three major saturation performance models have been proposed for
DCF; \textit{i)} assuming constant collision probability for each
station, Bianchi \cite{Bianchi00} developed a simple Discrete-Time
Markov Chain (DTMC) and the saturation throughput is obtained by
applying regenerative analysis to a generic slot time,
\textit{ii)} Cali \textit{et al.} \cite{Cali98},\cite{Cali00}
employed renewal theory to analyze a \textit{p}-persistent variant
of DCF with persistence factor \textit{p} derived from the CW, and
\textit{iii)} Tay \textit{et al.} \cite{Tay01} instead used an
average value mathematical method to model DCF backoff procedure
and to calculate the average number of interruptions that the
backoff timer experiences. Having the common assumption of slot
homogeneity (for an arbitrary station, constant collision or
transmission probability at an arbitrary slot), these models
define all different renewal cycles all of which lead to accurate
saturation performance analysis. Similarly, Medepalli \textit{et
al.} \cite{Medepalli05} provided explicit expressions for average
DCF cycle time and system throughput. Pointing out another
direction for future performance studies, Hui \textit{et al.}
\cite{Hui06} recently proposed the application of metamodeling
techniques in order to find approximate closed-form mathematical
models.

These major methods are modified by several researchers to include
the extra features of the EDCA function in the saturation
analysis. Xiao \cite{Xiao04},\cite{Xiao05} extended
\cite{Bianchi00} to analyze only the CW differentiation. Kong
\textit{et al.} \cite{Kong04} took AIFS differentiation into
account via a 3-dimensional DTMC. On the other hand, these EDCA
extensions miss the treatment of varying collision probabilities
at different AIFS slots due to varying number of contending
stations. Robinson \textit{et al.}
\cite{Robinson04},\cite{Robinson04_2} proposed an average analysis
on the collision probability for different contention zones during
AIFS and employed calculated average collision probability on a
2-dimensional DTMC.
Hui \textit{et al.} \cite{Hui04},\cite{Hui05} unified several
major approaches into one approximate average model taking into
account varying collision probability in different backoff
subperiods (corresponds to contention zones in \cite{Robinson04}).
Zhu \textit{et al.} \cite{Zhu05} proposed another analytical EDCA
Markov model averaging the transition probabilities based on the
number and the parameters of high priority flows. Inan \textit{et
al.} \cite{Inan07_ICC} proposed a simple DTMC which provides
accurate treatment of AIFS and CW differentiation between the ACs
for the constant transmission probability assumption. Another
3-dimensional DTMC is proposed by Tao \textit{et al.}
\cite{Tao04},\cite{Tao06} in which the third dimension models the
state of backoff slots between successive transmission periods. In
\cite{Tao04},\cite{Tao06}, the fact that the number of idle slots
between successive transmissions can be at most the minimum of
AC-specific $CW_{max}$ values is considered. Independent from
\cite{Tao04},\cite{Tao06}, Zhao \textit{et al.} \cite{Zhao02} had
previously proposed a similar model for the heterogeneous case
where each station has traffic of only one AC. Banchs \textit{et
al.} \cite{Banchs05},\cite{Banchs06} proposed another model which
considers varying collision probability among different AIFS slots
due to a variable number of stations. Chen \textit{et al.}
\cite{Chen03}, Kuo \textit{et al.} \cite{Kuo03}, and Lin
\textit{et al.} \cite{Lin06} extended \cite{Tay01} in order to
include mean value analysis for AIFS and CW differentiation.

Although it has not yet received much attraction, the research
that releases the saturation assumption basically follows two
major methods; \textit{i)} modeling the non-saturated behavior of
DCF or EDCA function via Markov analysis, \textit{ii)} employing
queueing theory \cite{Kleinrock75} and calculating certain
quantities through average or Markov analysis. Our approach in
this work falls into the first category.

Markov analysis for the non-saturated case still assumes slot
homogeneity and extends \cite{Bianchi00} with necessary extra
Markov states and transitions. Duffy \textit{et al.}
\cite{Duffy05} and Alizadeh-Shabdiz \textit{et al.}
\cite{Shabdiz04},\cite{Shabdiz06} proposed similar extensions of
\cite{Bianchi00} for non-saturated analysis of 802.11 DCF. Due to
specific structure of the proposed DTMCs, these extensions assume
a MAC layer buffer size of one packet. We show that this
assumption may lead to significant performance prediction errors
for EDCA in the case of larger buffers. Cantieni \textit{et al.}
\cite{Cantieni05} extended the model of \cite{Shabdiz04} assuming
infinitely large station buffers and the MAC queue being empty
with constant probability regardless of the backoff stage the
previous transmission took place. Li \textit{et al.} \cite{Li05}
proposed an approximate model for non-saturation where only CW
differentiation is considered. Engelstad \textit{et al.}
\cite{Engelstad06} used a DTMC model to perform delay analysis for
both DCF and EDCA considering queue utilization probability as in
\cite{Cantieni05}. Zaki \textit{et al.} \cite{Zaki04} proposed yet
another Markov model with states that are of fixed real-time
duration which cannot capture the pre-saturation DCF throughput
peak.

A number of models employing queueing theory have also been
developed for 802.11(e) performance analysis in non-saturated
conditions. These models are assisted by independent analysis for
the calculation of some quantities such as collision and
transmission probabilities. Tickoo \textit{et al.}
\cite{Tickoo04_1},\cite{Tickoo04_2} modeled each 802.11 node as a
discrete time G/G/1 queue to derive the service time distribution,
but the models are based on an assumption that the saturated
setting provides good approximation for certain quantities in
non-saturated conditions. Chen \textit{et al.} \cite{XChen06}
employed both G/M/1 and G/G/1 queue models on top of \cite{Xiao05}
which only considers CW differentiation. Lee \textit{et al.}
\cite{Lee06} analyzed the use of M/G/1 queueing model while
employing a simple non-saturated Markov model to calculate
necessary quantities. Medepalli \textit{et al.}
\cite{Medepalli05_2} built upon the average cycle time derivation
\cite{Medepalli05} to obtain individual queue delays using both
M/G/1 and G/G/1 queueing models. Foh \textit{et al.} \cite{Foh02}
proposed a Markov framework to analyze the performance of DCF
under statistical traffic. This framework models the number of
contending nodes as an M/E$_{j}$/1/k queue. Tantra \textit{et al.}
\cite{Tantra06} extended \cite{Foh02} to include service
differentiation in EDCA. However, such analysis is only valid for
a restricted scenario where all nodes have a MAC queue size of one
packet.


There are also a few studies that investigated the effect of EDCA
TXOPs on 802.11e performance for a saturated scenario. Mangold
\textit{et al.} \cite{Mangold02} and Suzuki \textit{et al.}
\cite{Suzuki06} carried out the performance analysis through
simulation. The efficiency of burst transmissions with block
acknowledgements is studied in \cite{Tinnirello05_2}. Tinnirello
\textit{et al.} \cite{Tinnirello05} also proposed different TXOP
managing policies for temporal fairness provisioning. Peng
\textit{et al.} \cite{Peng06} proposed an analytical model to
study the effect of burst transmissions and showed that improved
service differentiation can be achieved using a novel scheme based
on TXOP thresholds.

A thorough and careful literature survey shows that an EDCA
analytical model which incorporates \textit{all} EDCA QoS
parameters, CW, AIFS, and TXOP, for \textit{any} traffic load has
not been designed yet.

\section{EDCA Discrete-Time Markov Chain
Model}\label{sec:DTMCmodel}

Assuming slot homogeneity, we propose a novel DTMC to model the
behavior of the EDCA function of any AC at any load. The main
contribution of this work is that the proposed model considers the
effect of all EDCA QoS parameters (CW, AIFS, and TXOP) on the
performance for the whole traffic load range from a lightly-loaded
non-saturated channel to a heavily congested saturated medium.
Although we assume constant probability of packet arrival per
state (for the sake of simplicity, Poisson arrivals), we show that
the model provides accurate performance analysis for a range of
traffic types.

The state of the EDCA function of any AC at an arbitrary time $t$
depends on several MAC layer events that may have occured before
$t$. We model the MAC layer state of an AC$_{i}$, $0 \leq i \leq
3$, with a 3-dimensional Markov process,
$(s_{i}(t),b_{i}(t),q_{i}(t))$. The stochastic process $s_{i}(t)$
represents the value of the backoff stage at time $t$, i.e., the
number of retransmissions that the packet to be transmitted
currently has experienced until time $t$. The stochastic process
$b_{i}(t)$ represents the state of the backoff counter at time
$t$. Up to this point, the definition of the first two dimensions
follows \cite{Bianchi00} which is introduced for DCF. In order to
enable the accurate non-saturated analysis considering EDCA TXOPs,
we introduce another dimension which models the stochastic process
$q_{i}(t)$ denoting the number of packets buffered for
transmission at the MAC layer. Moreover, as the details will be
described in the sequel, in our model, $b_{i}(t)$ does not only
represent the value of the backoff counter, but also the number of
transmissions carried out during the current EDCA TXOP (when the
value of backoff counter is actually zero).

Using the assumption of independent and constant collision
probability at an arbitrary backoff slot, the 3-dimensional
process $(s_{i}(t),b_{i}(t),q_{i}(t))$ is represented as a
Discrete-Time Markov Chain (DTMC) with states $(j,k,l)$ and index
$i$. We define the limits on state variables as $0 \leq j \leq
r_{i}-1$, $-N_{i} \leq k \leq W_{i,j}$ and $0 \leq l \leq QS_{i}$.
In these inequalities, we let $r_{i}$ be the retransmission limit
of a packet of AC$_{i}$; $N_{i}$ be the maximum number of
successful packet exchange sequences of AC$_{i}$ that can fit into
one TXOP$_{i}$; $W_{i,j} = 2^{min(j,m_{i})}(CW_{i,min}+1)-1$ be
the CW size of AC$_{i}$ at the backoff stage $j$ where $CW_{i,max}
= 2^{m_{i}}(CW_{i,min}+1)-1$, $0\leq m_{i} < r_{i}$; and $QS_{i}$
be the maximum number of packets that can buffered at the MAC
layer, i.e., MAC queue size. Moreover, it is important to note
that a couple of restrictions apply to the state indices.
\begin{itemize}
\item When there are not any buffered packets at the AC queue, the
EDCA function of the corresponding AC cannot be in a
retransmitting state. Therefore, if $l=0$, then $j=0$ should hold.
Such backoff states represent the postbackoff process
\cite{802.11},\cite{802.11e}, therefore called as
\textit{postbackoff slots} in the sequel. The postbackoff
procedure ensures that the transmitting station waits at least
another backoff between successive TXOPs. Note that, when $l>0$
and $k \geq 0$, these states are named \textit{backoff slots}.
\item The states with indices $-N_{i}\leq k \leq -1$ represent the
negation of the number of packets that are successfully
transmitted at the current TXOP rather than the value of the
backoff counter (which is zero during a TXOP). For simplicity, in
the design of the Markov chain, we introduced such states in the
second dimension. Therefore, if $-N_{i}\leq k \leq -1$, we set
$j=0$. As it will be clear in the sequel, the addition of these
states enables EDCA TXOP analysis.
\end{itemize}

Let $p_{c_{i}}$ denote the average conditional probability that a
packet from AC$_{i}$ experiences either an external or an internal
collision after the EDCA function decides on the transmission. Let
$p_{nt}(l',T|l)$ be the probability that there are $l'$ packets in
the MAC buffer at time $t+T$ given that there were $l$ packets at
$t$ and no transmissions have been made during interval $T$.
Similarly, let $p_{st}(l',T|l)$ be the probability that there are
$l'$ packets in the MAC buffer at time $t+T$ given that there were
$l$ packets at time $t$ and a transmission has been made during
interval $T$. Note that since we assume Poisson arrivals, the
exponential interarrival distributions are independent, and
$p_{nt}$ and $p_{st}$ only depend on the interval length $T$ and
are independent of time $t$. Then, the nonzero state transmission
probabilities of the proposed Markov model for AC$_{i}$, denoted
as $P_{i}(j',k',l'|j,k,l)$ adopting the same notation in
\cite{Bianchi00}, are calculated as follows.
\begin{enumerate}
\item The backoff counter is decremented by one at the slot
boundary. Note that we define the postbackoff or the backoff slot
as Bianchi defines the slot time \cite{Bianchi00}. Then, for
$0\leq j \leq r_{i}-1$, $1\leq k \leq W_{i,j}$, and $0\leq l \leq
l' \leq QS_{i}$,
\begin{align}
\label{eq:unsatDTMC1} P_{i}(j,k-1,l'|j,k,l) =
p_{nt}(l',T_{i,bs}|l).
\end{align}

It is important to note that the proposed DTMC's evolution is not
real-time and the state duration varies depending on the state.
The average duration of a backoff slot $T_{i,bs}$ is calculated
by~(\ref{eq:unsatTbs}) which will be derived. Note that,
in~(\ref{eq:unsatDTMC1}), we consider the probability of packet
arrivals during $T_{i,bs}$ (buffer size $l'$ after the state
transition depends on this probability).

\item We assume the transmitted packet experiences a collision
with constant probability $p_{c_{i}}$ (slot homogeneity).
In the following, note that the cases when the retry limit is
reached and when the MAC buffer is full are treated separately,
since the transition probabilities should follow different rules.
Let $T_{i,s}$ and $T_{i,c}$ be the time spent in a successful
transmission and a collision by AC$_{i}$ respectively which will
be derived.
%
Then, for $0\leq j \leq r_{i}-1$, $0\leq l \leq QS_{i}-1$, and
$\max(0,l-1)\leq l' \leq QS_{i}$,
\begin{align}
\label{eq:unsatDTMC3} & P_{i}(0,-1,l'|j,0,l) = (1-p_{c_{i}})\cdot
p_{st}(l',T_{i,s}|l) \\
\label{eq:unsatDTMC4} & P_{i}(0,-1,QS_{i}-1|j,0,QS_{i}) =
1-p_{c_{i}}.
\end{align}

For $0\leq j \leq r_{i}-2$, $0\leq k \leq W_{i,j+1}$, and $0\leq l
\leq l' \leq QS_{i}$,
\begin{align}
\label{eq:unsatDTMC5} & P_{i}(j+1,k,l'|j,0,l) =
\frac{p_{c_{i}}\cdot p_{nt}(l',T_{i,c}|l)}{W_{i,j+1}+1}.
\end{align}

For $0\leq k \leq W_{i,0}$, $0\leq l \leq QS_{i}-1$, and
$\max(0,l-1)\leq l' \leq QS_{i}$,
\begin{align}
\label{eq:unsatDTMC6} & P_{i}(0,k,l'|r_{i}-1,0,l) =
\frac{p_{c_{i}}}{W_{i,0}+1}\cdot p_{st}(l',T_{i,s}|l) \\
\label{eq:unsatDTMC7} & P_{i}(0,k,QS_{i}-1|r_{i}-1,0,QS_{i}) =
\frac{p_{c_{i}}}{W_{i,0}+1}
\end{align}

Note that we use $p_{nt}$ in~(\ref{eq:unsatDTMC5}) although a
transmission has been made. On the other hand, the packet has
collided and is still at the MAC queue for retransmission as if no
transmission has occured. This is not the case in
(\ref{eq:unsatDTMC3}) and (\ref{eq:unsatDTMC6}), since in these
transitions a successful transmission or a drop occurs. When the
MAC buffer is full, any arriving packet is discarded as
(\ref{eq:unsatDTMC4}) and (\ref{eq:unsatDTMC7}) imply.


%

\item Once the TXOP is started, the EDCA function may continue
with as many packet SIFS-separated exchange sequences as it can
fit into the TXOP duration. Let $T_{i,exc}$ be the average
duration of a successful packet exchange sequence for AC$_{i}$
which will be derived in~(\ref{eq:unsatTexc}). Then, for $-N_{i}+1
\leq k \leq -1$, $1\leq l \leq QS_{i}$, and $\max(0,l-1)\leq l'
\leq QS_{i}$,
\begin{align}
\label{eq:unsatDTMC8} P_{i}(0,k-1,l'|0,k,l) =
p_{st}(l',T_{i,exc}|l).
\end{align}
When the next transmission cannot fit into the remaining TXOP, the
current TXOP is immediately concluded  and the unused portion of
the TXOP is returned. By design, our model includes maximum number
of packets that can fit into one TXOP. Then, for $0\leq k \leq
W_{i,0}$ and $1\leq l \leq QS_{i}$,
\begin{align}
\label{eq:unsatDTMC9} P_{i}(0,k,l|0,-N_{i},l) =
\frac{1}{W_{i,0}+1}.
\end{align}
The TXOP ends when the MAC queue is empty. Then, for $0\leq k'
\leq W_{i,0}$ and $-N_{i}\leq k \leq -1$,
\begin{align}
\label{eq:unsatDTMC10} P_{i}(0,k',0|0,k,0) = \frac{1}{W_{i,0}+1}.
\end{align}

Note that no time passes in (\ref{eq:unsatDTMC9}) and
(\ref{eq:unsatDTMC10}), so the definition of these states and
transitions is actually not necessary for accuracy. On the other
hand, they simplify the DTMC structure and symmetry.

\item If the queue is still empty when the postbackoff counter
reaches zero, the EDCA function enters the idle state until
another packet arrival. Note (0,0,0) also represents the idle
state.
We make two assumptions; \textit{i)} At most one packet may arrive
during $T_{slot}$ with constant probability $\rho_{i}$
(considering the fact that $T_{slot}$ is in the order of
microseconds, the probability that multiple packets can arrive in
this interval is very small), \textit{ii)} if the channel is idle
at the slot the packet arrives at an empty queue, the transmission
will be successful at AIFS completion without any backoff. The
latter assumption is due to the following reason. While the
probability of the channel becoming busy during AIFS or a
collision occuring for the transmission at AIFS is very small at a
lightly loaded scenario, the probability of a packet arrival to an
empty queue is very small at a highly loaded scenario. As observed
via simulations, these assumptions do not lead to any noticeable
changes in the results while simplifying the Markov chain
structure and symmetry.
Then, for $0\leq k \leq W_{i,0}$ and $1\leq l \leq
QS_{i}$,
\begin{align}
\label{eq:unsatDTMC11} & P_{i}(0,0,0|0,0,0) =
(1-p_{c_{i}})\cdot (1-\rho_{i}) + p_{c_{i}}\cdot p_{nt}(0,T_{i,b}|0), \\
\label{eq:unsatDTMC12} & P_{i}(0,k,l|0,0,0) =
\frac{p_{c_{i}}}{W_{i,0}+1}\cdot p_{nt}(l,T_{i,b}|0), \\
\label{eq:unsatDTMC13} & P_{i}(0,-1,l|0,0,0) =
(1-p_{c_{i}})\cdot\rho_{i}\cdot p_{nt}(l,T_{i,s}|0).
\end{align}

Let $T_{i,b}$ in (\ref{eq:unsatDTMC11}) and (\ref{eq:unsatDTMC12})
be the length of a backoff slot given it is not idle. Note that
actually a successful transmission occurs in the state transition
in~(\ref{eq:unsatDTMC13}). On the other hand, the transmitted
packet is not reflected in the initial queue size state which is
0. Therefore, $p_{nt}$ is used instead of $p_{st}$.
\end{enumerate}

\begin{figure}
\center {
\begin{tabular}{c}
\epsfig{file=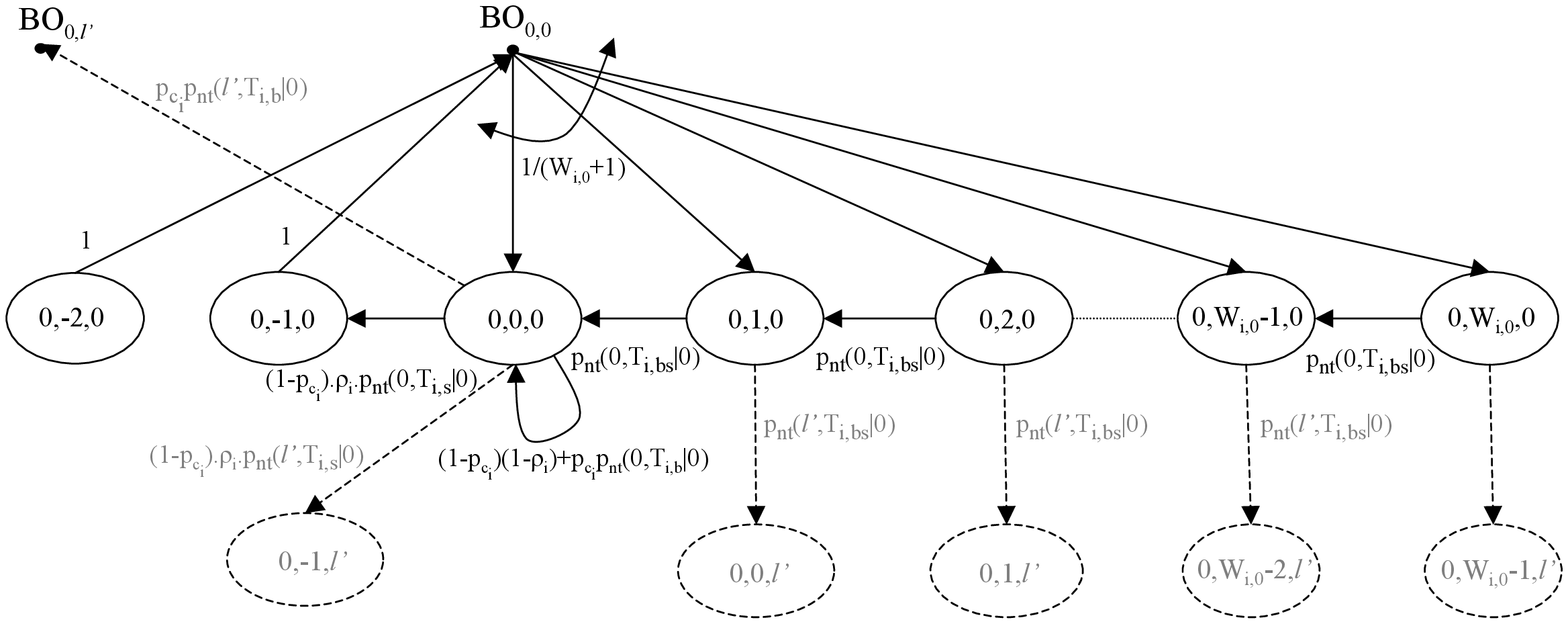,height=17 cm,angle=-90} \\ (a)
\\
\epsfig{file=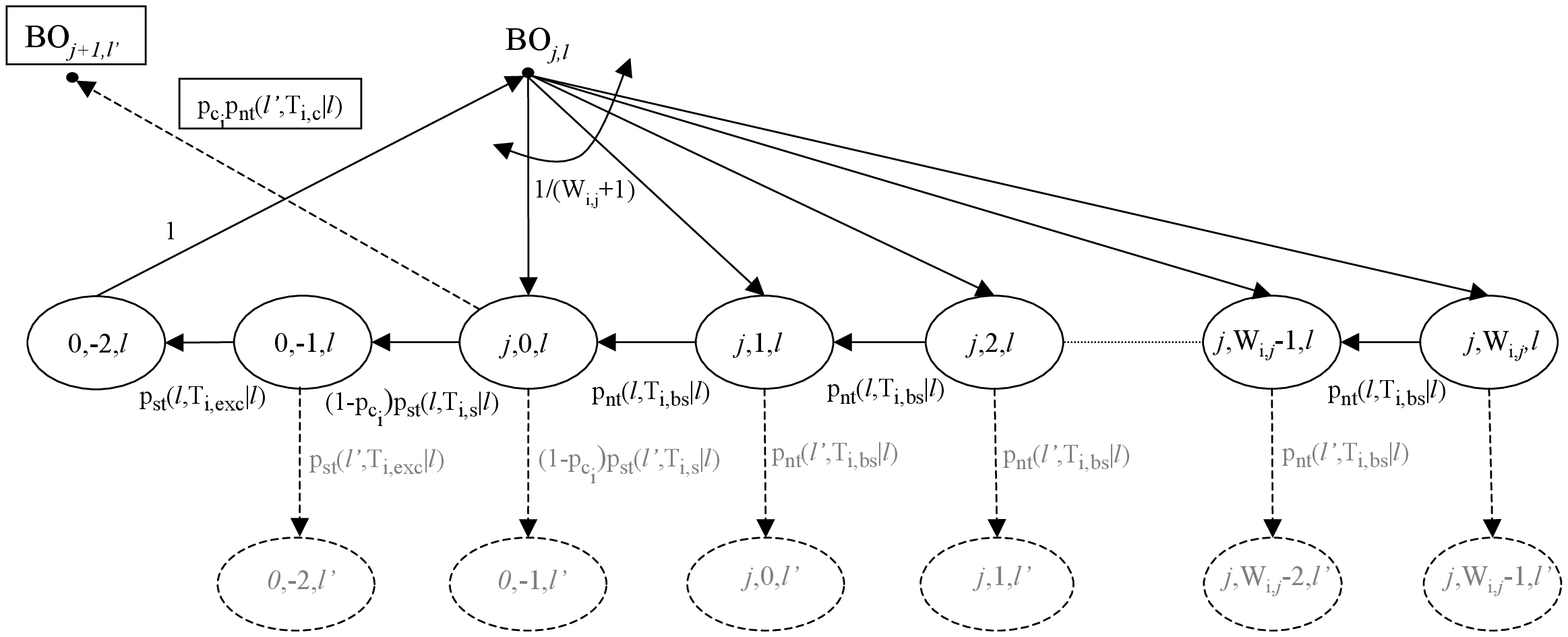,height=17 cm,angle=-90} \\ (b)
\\
\epsfig{file=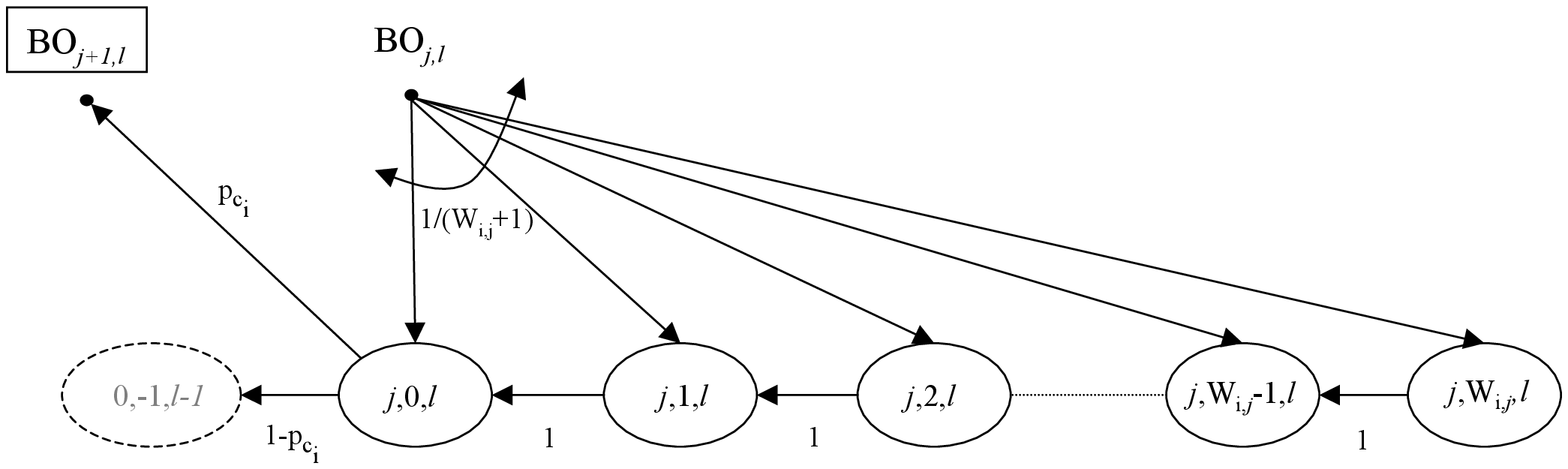,height=17 cm,angle=-90} \\ (c)
\\
\end{tabular}
} \caption{Parts of the proposed DTMC model for $N_{i}$=2. The
combination of these small chains for all $j$, $k$, $l$
constitutes the proposed DTMC model. (a) $l=0$. (b) $0<l<QS_{i}$.
(c) $l=QS_{i}$. Remarks: \textit{i)} the transition probabilities
and the states marked with rectangles differ when $j=r_{i}-1$ (as
in (\ref{eq:unsatDTMC6}) and (\ref{eq:unsatDTMC7})), \textit{ii)}
the limits for $l'$ follow the rules in
(\ref{eq:unsatDTMC1})-(\ref{eq:unsatDTMC13}).}
\label{fig:unsat_DTMCmodel}
\end{figure}

Parts of the proposed DTMC model are illustrated in
Fig.~\ref{fig:unsat_DTMCmodel} for an arbitrary AC$_{i}$ with
$N_{i}=2$. Fig.~\ref{fig:unsat_DTMCmodel}(a) shows the state
transitions for $l=0$. Note that in
Fig.~\ref{fig:unsat_DTMCmodel}(a) the states with $-N_{i}\leq k
\leq -2$ can only be reached from the states with $l=1$.
Fig.~\ref{fig:unsat_DTMCmodel}(b) presents the state transitions
for $0<l<QS_{i}$ and $0\leq j < r_{i}$. Note that only the
transition probabilities and the states marked with rectangles
differ when $j=r_{i}-1$ (as in (\ref{eq:unsatDTMC6})). Therefore,
we do not include an extra figure for this case.
Fig.~\ref{fig:unsat_DTMCmodel}(c) shows the state transitions when
$l=QS_{i}$. Note also that the states marked with rectangles
differ when $j=r_{i}-1$ (as in (\ref{eq:unsatDTMC7})). The
combination of these small chains for all $j$, $k$, $l$
constitutes our DTMC model.

\subsection{Steady-State Solution}

Let $b_{i,j,k,l}$ be the steady-state probability of the state
$(j,k,l)$ of the proposed DTMC with index $i$ which can be solved
using (\ref{eq:unsatDTMC1})-(\ref{eq:unsatDTMC13}) subject to
$\sum_{j}\sum_{k}\sum_{l}b_{i,j,k,l}=1$ (the proposed DTMC is
ergodic and irreducible). Let $\tau_{i}$ be the probability that
an AC$_{i}$ transmits at an arbitrary backoff or postbackoff slot
\begin{equation}
\label{eq:unsattau} \tau_{i}=
\frac{\left(\sum_{j=0}^{r_{i}-1}\sum_{l=1}^{QS_{i}}b_{i,j,0,l}\right)
+ b_{i,0,0,0} \cdot \rho_{i} \cdot
(1-p_{c_{i}})}{\sum_{j=0}^{r_{i}-1}\sum_{k=0}^{W_{i,j}}\sum_{l=0}^{QS_{i}}b_{i,j,k,l}}.
\end{equation}
\noindent Note that $-N_{i} \leq k \leq -1$ is not included in the
normalization in (\ref{eq:unsattau}), since these states represent
a continuation in the EDCA TXOP rather than a contention for the
access. The value of $\tau_{i}$ depends on the values of the
average conditional collision probability $p_{c_{i}}$, the various
state durations $T_{i,bs}$, $T_{i,b}$, $T_{i,s}$ and $T_{i,c}$,
and the conditional queue state transition probabilities $p_{nt}$
and $p_{st}$.

\subsubsection{Average conditional collision probability
$p_{c_{i}}$}

\begin{figure}
\center{\epsfig{file=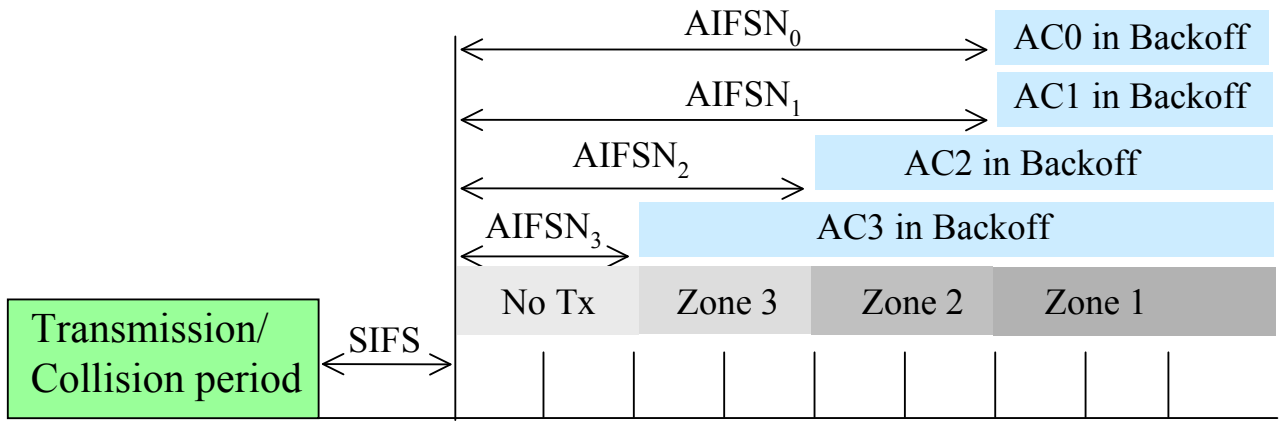,height=17
cm,angle=-90}} \caption[] {\label{fig:unsat_contzones} EDCA
backoff after busy medium.
}
\end{figure}

The difference in AIFS of each AC in EDCA creates the so-called
\textit{contention zones} as shown in
Fig.~\ref{fig:unsat_contzones} \cite{Robinson04}. In each
contention zone, the number of contending stations may vary. The
collision probability cannot simply be assumed to be constant
among all ACs.

We can define $p_{c_{i,x}}$ as the conditional probability that
AC$_{i}$ experiences either an external or an internal collision
given that it has observed the medium idle for $AIFS_{x}$ and
transmits in the current slot (note $AIFS_{x}\geq AIFS_{i}$ should
hold). For the following, in order to be consistent with the
notation of \cite{802.11e}, we assume $AIFS_{0}\geq AIFS_{1} \geq
AIFS_{2} \geq AIFS_{3}$. Let $d_{i} = AIFS_{i} - AIFS_{3}$. Also,
let the total number AC$_{i}$ flows be $f_{i}$. Then, for the
heterogeneous scenario in which each station has only one AC
\begin{equation}
\label{eq:unsatpcix} \setlength{\nulldelimiterspace}{0pt}
p_{c_{i,x}} = 1-\frac{\prod\limits_{i':d_{i'}\leq d_{x}}
(1-\tau_{i'})^{f_{i'}}}{(1-\tau_{i})}.
\end{equation}

When each station has multiple ACs that are active, internal
collisions may occur. Then, for the scenario in which each station
has all 4 ACs active
\begin{equation}
\label{eq:unsatpcixmul} p_{c_{i,x}} = 1-\prod_{i':d_{i'}\leq
d_{x}} (1-\tau_{i'})^{f_{i'}-1}\prod_{i''>i}(1-\tau_{i''}).
\end{equation}

\noindent Similar extensions when the number of active ACs are 2
or 3 are straightforward.

\begin{figure}
\center{\epsfig{file=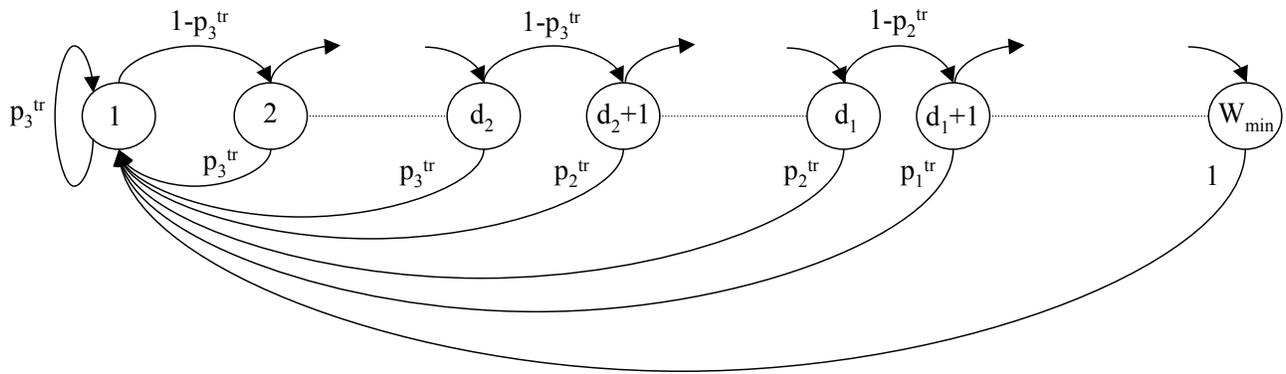,height=17
cm,angle=-90}} \caption[] {\label{fig:unsat_AIFSMC} Transition
through backoff slots in different contention zones for the
example given in Fig.\ref{fig:unsat_contzones}.}
\end{figure}

We use the Markov chain shown in Fig.~\ref{fig:unsat_AIFSMC} to
find the long term occupancy of contention zones. Each state
represents the $n^{th}$ backoff slot after completion of the
AIFS$_{3}$ idle interval following a transmission period. The
Markov chain model uses the fact that a backoff slot is reached if
and only if no transmission occurs in the previous slot. Moreover,
the number of states is limited by the maximum idle time between
two successive transmissions which is $W_{min}=\min(CW_{i,max})$
for a saturated scenario. Although this is not the case for a
non-saturated scenario, we do not change this limit. As the
comparison with simulation results show, this approximation does
not result in significant prediction errors. The probability that
at least one transmission occurs in a backoff slot in contention
zone $x$ is
\begin{equation}
\label{eq:unsatptr} \setlength{\nulldelimiterspace}{0pt}
p^{tr}_{x} = 1-\prod_{i':d_{i'}\leq d_{x}} (1-\tau_{i'})^{f_{i'}}.
\end{equation}

\noindent Note that the contention zones are labeled with $x$
regarding the indices of $d$. In the case of equal AIFS values,
the contention zone is labeled with the index of the AC with
higher priority.

Given the state transition probabilities as in
Fig.~\ref{fig:unsat_AIFSMC}, the long term occupancy of the
backoff slots $b'_{n}$ can be obtained from the steady-state
solution of the Markov chain. Then, the AC-specific average
collision probability $p_{c_{i}}$ is found by weighing zone
specific collision probabilities $p_{c_{i,x}}$ according to the
long term occupancy of contention zones (thus backoff slots)
\begin{equation}
\label{eq:unsatpci}p_{c_{i}} = \frac{\sum_{n=d_{i}+1}^{W_{min}}
p_{c_{i,x}}\cdot b'_{n}}{\sum_{n=d_{i}+1}^{W_{min}} b'_{n}}
\end{equation}
\noindent where $x = \max \left( y~|~d_{y} = \underset{z}{\max}
(d_{z}~|~d_{z} \leq n)\right)$ which shows $x$ is assigned the
highest index value within a set of ACs that have AIFS smaller
than or equal to $n+AIFS_{3}$. This ensures that at backoff slot
$n$, AC$_{i}$ has sensed the medium idle for AIFS$_{x}$.
Therefore, the calculation in~(\ref{eq:unsatpci}) fits into the
definition of $p_{c_{i,x}}$.

Note that the average collision probability calculation in
\cite[Section IV-D]{Robinson04} is a special case of our
calculation for two ACs.

\subsubsection{The state duration $T_{i,s}$ and $T_{i,c}$}

Let $T_{i,p}$ be the average payload transmission time for
AC$_{i}$ ($T_{i,p}$ includes the transmission time of MAC and PHY
headers), $\delta$ be the propagation delay, $T_{ack}$ be the time
required for acknowledgment packet (ACK) transmission. Then, for
the basic access scheme, we define the time spent in a successful
transmission $T_{i,s}$ and a collision $T_{i,c}$ for any AC$_{i}$
as
\begin{align}\label{eq:unsatTs}
T_{i,s} = & T_{i,p} + \delta + SIFS + T_{ack} + \delta + AIFS_{i}
\\ \label{eq:unsatTc} T_{i,c} = & T_{i,p^{*}} + ACK\_Timeout +
AIFS_{i}
\end{align}
\noindent where $T_{i,p^{*}}$ is the average transmission time of
the longest packet payload involved in a collision
\cite{Bianchi00}. For simplicity, we assume the packet size to be
equal for any AC, then $T_{i,p^{*}}=T_{i,p}$. Being not explicitly
specified in the standards, we set $ACK\_Timeout$, using Extended
Inter Frame Space (EIFS) as $EIFS_{i}-AIFS_{i}$.

The extensions of~(\ref{eq:unsatTs}) and~(\ref{eq:unsatTc}) for
the Request-to-Send/Clear-to-Send (RTS/CTS) scheme are
\begin{align}\label{eq:unsatTs_RTSCTS}
T_{i,s} = & T_{rts} + \delta + SIFS + T_{cts} + \delta + SIFS +
T_{i,p} + \delta + SIFS + T_{ack} + \delta + AIFS_{i}
\\ \label{eq:unsatTc_RTSCTS} T_{i,c} = & T_{rts} + CTS\_Timeout +
AIFS_{i}
\end{align}
\noindent where $T_{rts}$ and $T_{cts}$ are the time required for
RTS and CTS packet transmissions respectively. Being not
explicitly specified in the standards, we set $CTS\_Timeout$ as we
set $ACK\_Timeout$.

\subsubsection{The state duration $T_{i,bs}$ and $T_{i,b}$}

The average time between successive backoff counter decrements is
denoted by $T_{i,bs}$. The backoff counter decrement may be at the
slot boundary of an idle backoff slot or the last slot of AIFS
following an EDCA TXOP or a collision period. We start with
calculating the average duration of an EDCA TXOP for AC$_{i}$
$T_{i,txop}$ as
\begin{equation}
\label{equnsatTtxop} T_{i,txop} = \frac{
\sum_{l=0}^{QS_{i}}b_{i,0,-N_{i},l}\cdot ((N_{i}-1)\cdot T_{i,exc}
+ T_{i,s}) + \sum_{k=-N_{i}+1}^{-1}b_{i,0,k,0}\cdot ((-k-1)\cdot
T_{i,exc} +
T_{i,s})}{\sum_{k=-N_{i}+1}^{-1}b_{i,0,k,0}+\sum_{l=0}^{QS_{i}}b_{i,0,-N_{i},l}}
\end{equation}
\noindent where $T_{i,exc}$ is defined as the duration of a
successful packet exchange sequence within a TXOP. Since the
packet exchanges within a TXOP are separated by SIFS rather than
AIFS,
\begin{align}
\label{eq:unsatTexc} T_{i,exc} = T_{i,s} - AIFS_{i} + SIFS,
\\
\label{eq:unsatNi} N_{i} = \max (1,\lfloor
(TXOP_{i}+SIFS)/T_{i,exc} \rfloor) .
\end{align}

Given $\tau_{i}$ and $f_{i}$, simple probability theory can be
used to calculate the conditional probability of no transmission
($p_{x,i}^{idle}$), only one transmission from AC$_{i'}$
($p_{x,i}^{suc_{i'}}$), or at least two transmissions
($p_{x,i}^{col}$) at the contention zone $x$ given one AC$_{i}$ is
in backoff.
\begin{equation}
\label{eq:unsatnotrans} \setlength{\nulldelimiterspace}{0pt}
p_{x,i}^{idle} =
\left\{ \\
\begin{IEEEeqnarraybox}[\relax][c]{ll} \prod_{i':d_{i'}\leq d_{x}}(1-\tau_{i'})^{f_{i'}}, & ~ {\rm if}~d_{i} > d_{x} \\
\frac{\prod_{i':d_{i'}\leq
d_{x}}(1-\tau_{i'})^{f_{i'}}}{1-\tau_{i}}, & ~ {\rm if} ~d_{i}
\leq d_{x}.
\end{IEEEeqnarraybox}
\right.
\end{equation}
\begin{equation}
\label{eq:unsatonetrans} \setlength{\nulldelimiterspace}{0pt}
p_{x,i}^{suc_{i'}} =
\left\{ \\
\begin{IEEEeqnarraybox}[\relax][c]{ll} 0, & ~ {\rm if} ~d_{x}
< d_{i'} \\
f_{i'}\tau_{i'}(1-\tau_{i'})^{f_{i'}-1}\prod_{i'':d_{i''}\leq
d_{x}}(1-\tau_{i''})^{f_{i''}}, & ~ {\rm if}~d_{i} >
d_{x}~{\rm and}~d_{i'} \leq d_{x} \\
\frac{f_{i'}\tau_{i'}(1-\tau_{i'})^{f_{i'}-1}}{1-\tau_{i}}\prod_{i'':d_{i''}\leq
d_{x}}(1-\tau_{i''})^{f_{i''}}, & ~ {\rm if} ~d_{i} \leq
d_{x}~{\rm and}~d_{i'} \leq d_{x}.
\end{IEEEeqnarraybox}
\right.
\end{equation}
\begin{align}
\label{eq:unsatmulttrans} p_{x,i}^{col} = & 1 - p_{x,i}^{idle} -
\sum_{\forall i'} p_{x,i}^{suc_{i'}}
\end{align}

Let $x_{i}$ be the first contention zone in which AC$_{i}$ can
transmit. Then,
\begin{align}
\label{eq:unsatTbs} T_{i,bs} = \frac{1}{1-\sum\limits_{x_{i} < x'
\leq 3}p_{z_{x'}}}\sum_{\forall x'}(p_{x',i}^{idle}\cdot
T_{slot}+p_{x',i}^{col}\cdot T_{i,c}+\sum_{\forall
i'}p_{x',i}^{suc_{i'}}\cdot T_{i',txop})\cdot p_{z_{x'}}
\end{align}
\noindent where $p_{z_{x}}$ denotes the stationary distribution
for a random backoff slot being in zone $x$. Note that, in
(\ref{eq:unsatTbs}), the fractional term before summation accounts
for the busy periods experienced before AIFS$_{i}$ is completed.
Therefore, if we let $d_{-1}=W_{min}$,
\begin{align}
\label{eq:unsatpzx} p_{z_{x}} =
\sum_{n=d_{x}+1}^{min(d_{x'}|d_{x'}>d_{x})}b_{n}'.
\end{align}

The expected duration of a backoff slot given it is busy and one
AC$_{i}$ is in idle state is calculated as
\begin{align}
\label{eq:unsatTb} T_{i,b} = \sum_{\forall
x'}\left(\frac{p_{x',i}^{col}}{1-p_{x',i}^{idle}}\cdot
T_{i,c}+\sum_{\forall
i'}\frac{p_{x',i}^{suc_{i'}}}{1-p_{x',i}^{idle}}\cdot
T_{i',txop}\right)\cdot p_{z_{x'}}.
\end{align}

\subsubsection{The conditional queue state transition probabilities $p_{nt}$
and $p_{st}$} We  assume the packets arrive at the AC queue with
size $QS_{i}$ according to a Poisson process with rate
$\lambda_{i}$ packets per second. Using the probability
distribution function of the Poisson process, the probability of
$k$ arrivals occuring in time interval $t$ can be calculated as
\begin{align}
{\rm Pr}(N_{t,i}=k) = \frac{\exp^{-\lambda_{i} t}(\lambda_{i}
t)^{k}}{k!}.
\end{align}

Then, $p_{nt}(l',T|l)$ and $p_{st}(l',T|l)$ can be calculated as
follows. Note that the finite buffer space is considered
throughout calculations since the number of packets that may
arrive during $T$ can be more than the available queue space.
\begin{equation}
\label{eq:unsatpnt} \setlength{\nulldelimiterspace}{0pt}
p_{nt}(l',T|l) =
\left\{ \\
\begin{IEEEeqnarraybox}[\relax][c]{ll} {\rm Pr}(N_{T,i}=l'-l), & ~ {\rm if}~l' < QS_{i} \\
1 - \sum_{l'=l}^{QS_{i}-1}{\rm Pr}(N_{T,i}=l'-l), & ~ {\rm if} ~
l'=QS_{i}.
\end{IEEEeqnarraybox}
\right.
\end{equation}
\begin{equation}
\label{eq:unsatpst} \setlength{\nulldelimiterspace}{0pt}
p_{st}(l',T|l) =
\left\{ \\
\begin{IEEEeqnarraybox}[\relax][c]{ll} {\rm Pr}(N_{T,i}=l'-l+1), & ~ {\rm if}~l' < QS_{i} \\
1 - \sum_{l'=l-1}^{QS_{i}-1}{\rm Pr}(N_{T,i}=l'-l+1), & ~ {\rm if}
~ l'=QS_{i}.
\end{IEEEeqnarraybox}
\right.
\end{equation}

Note that in (\ref{eq:unsatDTMC11})-(\ref{eq:unsattau}), $\rho_{i}
= 1-{\rm Pr}(N_{T_{slot},i}=0)$. Together with the steady-state
transition probabilities, (\ref{eq:unsattau})-(\ref{eq:unsatpst})
represent a nonlinear system which can be solved using numerical
methods.

\subsection{Normalized Throughput Analysis}
The normalized throughput of a given AC$_{i}$, $S_{i}$, is defined
as the fraction of the time occupied by the successfully
transmitted information. Then,
\begin{align}
\label{eq:unsatSi_def} S_{i} = &
\frac{p_{s_{i}}N_{i,txop}T_{i,p}}{p_{I}T_{slot}+\sum_{i'}p_{s_{i'}}T_{i',txop}+(1-p_{I}-\sum_{i'}p_{s_{i'}})T_{c}}
\end{align}

\noindent $p_{I}$ is the probability of the channel being idle at
a backoff slot, $p_{s_{i}}$ is the conditional successful
transmission probability of AC$_{i}$ at a  backoff slot, and
$N_{i,txop} = (T_{i,txop}-AIFS_{i}+SIFS)/T_{i,exc}$. Note that, we
consider $N_{i,txop}$ and $T_{i,txop}$ in (\ref{eq:unsatSi_def})
to define the generic slot time and the time occupied by the
successfully transmitted information in the case of EDCA TXOPs.

The probability of a slot being idle, $p_{I}$, depends on the
state of previous slots. For example, conditioned on the previous
slot to be busy ($p_{B}=1-p_{I}$), $p_{I}$ only depends on the
transmission probability of the ACs with the smallest AIFS, since
others have to wait extra AIFS slots. Generalizing this to all
AIFS slots, $p_{I}$ can be calculated as
\begin{align}
\label{eq:unsatp_I} p_{I} & =
\sum_{n=0}^{W_{min}}\gamma_{n}p_{B}(p_{I})^{n} \cong
\sum_{n=0}^{d_{0}-1}\gamma_{n}p_{B}p_{I}^{n}+\gamma_{d_{0}}p_{I}^{d_{0}}
\end{align}

\noindent where $\gamma_{n}$ denotes the probability of no
transmission occuring at the $(n+1)^{th}$ AIFS slot after
$AIFS_{3}$. Substituting $\gamma_{n}=\gamma_{d_{0}}$ for $n\geq
d_{0}$, and releasing the condition on the upper limit of
summation, $W_{min}$, to $\infty$, $p_{I}$ can be approximated as
in~(\ref{eq:unsatp_I}). According to the simulation results, this
approximation works well. Note that $\gamma_{n} = 1-p_{x}^{tr}$
where $x = \max \left( y~|~d_{y} = \underset{z}{\max}
(d_{z}~|~d_{z} \leq n)\right)$.


The probability of successful transmission $p_{s_{i}}$ is
conditioned on the states of the previous slots as well. This is
again because the number of stations that can contend at an
arbitrary backoff slot differs depending on the number of previous
consecutive idle backoff slots. Therefore, for the heterogeneous
case, in which each station only has one AC, $p_{s_{i}}$ can be
calculated as
\begin{align}
\label{eq:unsatp_s_i} p_{s_{i}} =
\frac{N_{i}\tau_{i}}{(1-\tau_{i})} \left(
\sum_{n=d_{i}+1}^{d_{0}}\left(p_{B}p_{I}^{(n-1)}\prod_{i':0\leq
d_{i'}\leq (n-1)}(1-\tau_{i'})^{f_{i'}}\right) +
(p_{I})^{d_{0}}\prod_{\forall i'}(1-\tau_{i'})^{f_{i'}}\right).
\end{align}

Similarly, for the scenario, in which each station has four active
ACs,
\begin{align}
\label{eq:unsat_p_s_i_mul} p_{s_{i}} = &
\frac{N_{i}\tau_{i}}{(1-\tau_{i})} \left( \sum_{n=d_{i}+1}^{d_{0}}
\left( p_{B}p_{I}^{(n-1)}\prod_{i':0\leq d_{i'}\leq
(n-1)}(1-\tau_{i'})^{f_{i'}-1}\prod_{i''>i}(1-\tau_{i''})\right)
\right. \nonumber \\ & +  \left. (p_{I})^{d_{0}}\prod_{\forall
i'}(1-\tau_{i'})^{f_{i'}-1}\prod_{i''>i}(1-\tau_{i''})\right).
\end{align}

\subsection{Average Delay Analysis}

Our goal is to find total average delay $E[D_{i}]$ which is
defined as the average time from when a packet enters the MAC
layer queue of AC$_{i}$ until it is successfully transmitted.
$D_{i}$ has two components; \textit{i)} queueing time $Q_{i}$ and
\textit{ii)} access time $A_{i}$. $Q_{i}$ is the period that a
packet waits in the queue for other packets in front to be
transmitted. $A_{i}$ is the period a packet waits at the head of
the queue until it is transmitted successfully (backoff and
transmission period). We carry out a recursive calculation as in
\cite{Kong04} to find $E[A_{i}]$ for AC$_{i}$. Then, using
$E[A_{i}]$ and $b_{i,j,k,l}$, we calculate
$E[D_{i}]$=$E[Q_{i}]$+$E[A_{i}]$. Note that, $E[A_{i}]$ differs
depending on whether the EDCA function is idle or not when the
packet arrives. We will treat these cases separately. In the
sequel, $A_{i,idle}$ denotes the access delay when the EDCA
function is idle at the time a packet arrives.

The recursive calculation is carried out in a bottom-to-top and
left-to-right manner on the AC-specific DTMC. For the analysis,
let $A_{i}(j,k)$ denote the time delay from the current state
$(j,k,l)$ until the packet at the head of the AC$_{i}$ queue is
transmitted successfully ($l\geq 1$). The initial condition on the
recursive calculation is
\begin{equation}
\label{eq:unsatDinitial} \setlength{\nulldelimiterspace}{0pt}
A_{i}(r_{i}-1,0) = T_{i,s}.
\end{equation}
Recursive delay calculations for $0\leq j \leq r_{i}-1$ are
\begin{equation}
\label{eq:unsatDrec} \setlength{\nulldelimiterspace}{0pt}
A_{i}(j,k) =
\left\{ \\
\begin{array}[c]{ll} A_{i}(j,k-1) + T_{i,bs}, & ~ {\rm if}~1\leq k \leq W_{i,j} \\
(1-p_{c_{i}})
T_{i,s}+p_{c_{i}}\left(\frac{\sum_{k'=0}^{W_{i,j+1}}A_{i}(j+1,k')}{W_{i,j+1}+1}+T_{i,c}\right),
& ~{\rm if} ~ k = 0~{\rm and}~j\neq r_{i}-1.
\end{array}
\right.
\end{equation}
Then,
\begin{align}
\label{eq:unsatTad} E[A_{i}] =
\frac{\sum_{k=0}^{W_{i,0}}A_{i}(0,k)}{W_{i,0}+1}
\end{align}

Following the assumptions made in
(\ref{eq:unsatDTMC11})-(\ref{eq:unsatDTMC13}) and considering the
packet loss probability due to the retry limit as $p_{l,r} =
(p_{c_{i}})^{r_{i}}$ (note that the delay a dropped packet
experiences cannot be considered in a total delay calculation),
$E[A_{i,idle}]$ can be calculated as
\begin{equation}
E[A_{i,idle}] = T_{i,s} \cdot (1-p_{c_{i}}) + (E[A_{i}]+T_{i,b})
\cdot p_{c_{i}} \cdot (1-p_{l,r}).
\end{equation}

\noindent In this case, the average access delay is equal to the
total average delay, i.e., $D_{i}(0,0,0) = E[A_{i,idle}]$.

We perform another recursive calculation to calculate the total
delay a packet experiences $D_{i}(j,k,l)$ (given that the packet
arrives while the EDCA function is at state $(j,k,l)$). In the
calculations, we account for the remaining access delay for the
packet at the head of the MAC queue and the probability that this
packet may be dropped due to the retry limit.

Let $A_{i,d}(j,k)$ be the access delay conditioned that the packet
drops. $A_{i,d}(j,k)$ can easily be calculated by modifying the
recursive method of calculating $A_{i}(j,k)$. The initial
condition on this recursive calculation is
\begin{equation}
\label{eq:unsatDinitial} \setlength{\nulldelimiterspace}{0pt}
A_{i,d}(r_{i}-1,0) = T_{i,c}.
\end{equation}
Recursive delay calculations for $0\leq j \leq r_{i}-1$ are
\begin{equation}
\label{eq:unsatDrec} \setlength{\nulldelimiterspace}{0pt}
A_{i,d}(j,k) =
\left\{ \\
\begin{IEEEeqnarraybox}[\relax][c]{ll} A_{i}(j,k-1) + T_{i,bs}, & ~ {\rm if}~1\leq k \leq W_{i,j} \\
\sum_{k'=0}^{W_{i,j+1}}A_{i,d}(j+1,k)+T_{i,c}, & ~{\rm if} ~ k =
0~{\rm and}~j\neq r_{i}-1.
\end{IEEEeqnarraybox}
\right.
\end{equation}
Then,
\begin{align}
\label{eq:unsatTad} E[A_{i,d}] =
\frac{\sum_{k=0}^{W_{i,0}}A_{i,d}(0,k)}{W_{i,0}+1}
\end{align}

If a packet arrives during the backoff of another packet, it is
delayed at least for the remaining access time. Depending on the
queue size, it may be transmitted at the current TXOP, or may be
delayed till further accesses are gained. Then, for $0\leq j \leq
r_{i}-1$, $0\leq k \leq W_{i,j}$, and $1\leq l \leq QS_{i}$,
\begin{align}
D_{i}(j,k,l) = & (1-p_{l,r})\cdot \left(A_{i}(j,k) +
\min(N_{i}-1,l-1)\cdot T_{i,exc} + D_{i}(-1,-1,l-N_{i})\right) \nonumber \\
& + p_{l,r}\cdot \left(A_{i,d}(j,k) + D_{i}(-1,-1,l-1)\right).
\label{eq:unsat_e2edelay1}
\end{align}

\noindent When the packet arrives during postbackoff, the total
delay is equal to the access delay. Then, for $0\leq k \leq
W_{i,j}$ and $l=0$,
\begin{align}
D_{i}(j,k,l) = A_{i}(j,k). \label{eq:unsat_e2edelay2}
\end{align}

\noindent When the packet arrives during a TXOP, it may be
transmitted at the current TXOP, or it may wait for further
accesses. Then, for $-N_{i}+1\leq k \leq -1$ and $1\leq l \leq
QS_{i}$,
\begin{align}
D_{i}(j,k,l) = \min(k-1,l)\cdot T_{i,exc} + D_{i}(-1,-1,l-k+1).
\label{eq:unsat_e2edelay3}
\end{align}

$D_{i}(-1,-1,l)$ is calculated recursively according to the value
of $l$

\begin{equation}
\label{eq:unsat_e2edelay4} \setlength{\nulldelimiterspace}{0pt}
D_{i}(-1,-1,l) =
\left\{ \\
\begin{array}[c]{ll} 0, & ~ {\rm if}~l\leq0 \\
E[A_{i}] \cdot (1-p_{l,r}), & ~{\rm if} ~ l = 1 \\ \chi, &  ~ {\rm
if}~l>1
\end{array}
\right.
\end{equation}
\noindent where
\begin{align}
\chi = & (1-p_{l,r})\cdot \left(E[A_{i}] + \min(N_{i}-1,l-1)\cdot
T_{i,exc} \right. \nonumber \\ & \left. +
D_{i}(-1,-1,l-N_{i})\right) + p_{l,r}\cdot \left(E[A_{i,d}] +
D_{i}(-1,-1,l-1)\right). \label{eq:unsat_e2edelay4}
\end{align}

Let the probability of any arriving packet seeing the EDCA
function at state $(j,k,l)$ be $\bar{b}_{i,j,k,l}$. Since we
assume independent and exponentially distributed packet
interarrivals, $\bar{b}_{i,j,k,l}$ can simply be calculated by
normalizing $b_{i,j,k,l}$ excluding the states in which no time
passes, i.e., $\forall(j,k,l)$ such that $(0, -N_{i}, 1\leq l \leq
QS_{i})$ or $(0,-N_{i}\leq k \leq -1,0)$. Note that
$\bar{b}_{i,j,k,l}$ is zero for these states
\begin{equation}
\bar{b}_{i,j,k,l} =
\frac{b_{i,j,k,l}}{1-\sum_{l=1}^{QS_{i}}b_{i,0,-N_{i},l}-\sum_{k=-N_{i}}^{-1}b_{i,0,k,0}}.
\label{eq:unsat_bnormalized}
\end{equation}

\noindent Then, the total average delay a successful packet
experiences $E[D_{i}]$ can be calculated averaging $D_{i}(j,k,l)$
over all possible states
\begin{align}
E[D_{i}] = E[A_{i,idle}]\cdot \bar{b}_{i,0,0,0} + \sum_{\forall
(j,k,l)/(0,0,0)} D_{i}(j,k,l)\cdot
\bar{b}_{i,j,k,l}.\label{eq:unsat_e2edelay5}
\end{align}

\subsection{Average Packet Loss Ratio}

We consider two types of packet losses; \textit{i)} the packet is
dropped when the MAC layer retry limit is reached, \textit{ii)}
the packet is dropped if the MAC queue is full at the time of
packet arrival. Let $plr_{i}$ denote the average packet loss ratio
for AC$_{i}$. We use the steady-state probability $b_{i,j,k,l}$ to
find the probability whether the MAC queue is full or not at the
time of packet arrival. If the queue is full, the arriving packet
is dropped (second term in (\ref{eq:unsat_ave_plr})). Otherwise,
the packet is dropped with probability $p_{c_{i}}^{r_{i}}$, i.e.
only if the retry limit is reached (first term in
(\ref{eq:unsat_ave_plr})). Note that we consider packet
retransmissions only due to packet collisions. Then,
\begin{equation}
\label{eq:unsat_ave_plr} plr_{i} =
\sum_{j=0}^{r_{i}-1}\sum_{k=0}^{W_{i,j}}\sum_{l=0}^{QS_{i}-1}b_{i,j,k,l}\cdot
p_{c_{i}}^{r_{i}} +
\sum_{j=0}^{r_{i}-1}\sum_{k=0}^{W_{i,j}}b_{i,j,k,QS_{i}}.
\end{equation}

\subsection{Queue Size Distribution}

Due to the specific structure of the proposed model, it is
straightforward to calculate the MAC queue size distribution for
$AC_{i}$. Note that we use queue size distribution in the
calculation of average packet loss ratio.
\begin{equation}
\label{eq:unsat_ave_qsd} {\rm Pr}(l=l') =
\sum_{j=0}^{r_{i}-1}\sum_{k=0}^{W_{i,j}}b_{i,j,k,l'}.
\end{equation}

\section{Numerical and Simulation Results}\label{sec:simulations}

We validate the accuracy of the numerical results calculated via
the proposed EDCA model by comparing them with the simulations
results obtained from ns-2 \cite{ns2}. For the simulations, we
employ the IEEE 802.11e HCF MAC simulation model for ns-2.28 that
we developed \cite{ourcode}. This module implements all the EDCA
and HCCA functionalities stated in \cite{802.11e}.

As in all work on the subject in the literature, we consider ACs
that transmit fixed-size User Datagram Protocol (UDP) packets. In
simulations, we consider two ACs, one high priority and one low
priority. Each station runs only one traffic class. Unless
otherwise stated, the packets are generated according to a Poisson
process with equal rate for both ACs. We set $AIFSN_{1}=3$,
$AIFSN_{3}=2$, $CW_{1,min}=15$, $CW_{3,min}=7$, $m_{1}=m_{3}=3$,
$r_{1}=r_{3}=7$. For both ACs, the payload size is 1034 bytes.
Again, as in most of the work on the subject, the simulation
results are reported for the wireless channel which is assumed to
be not prone to any errors during transmission. The errored
channel case is left for future study. All the stations have
802.11g Physical Layer (PHY) using 54 Mbps and 6 Mbps as the data
and basic rate respectively ($T_{slot}=9~\mu s$, $SIFS=10~\mu s$)
\cite{802.11g}. The simulation runtime is 100 seconds.

\begin{figure*}[p]
\centering \includegraphics[width =
1.0\linewidth]{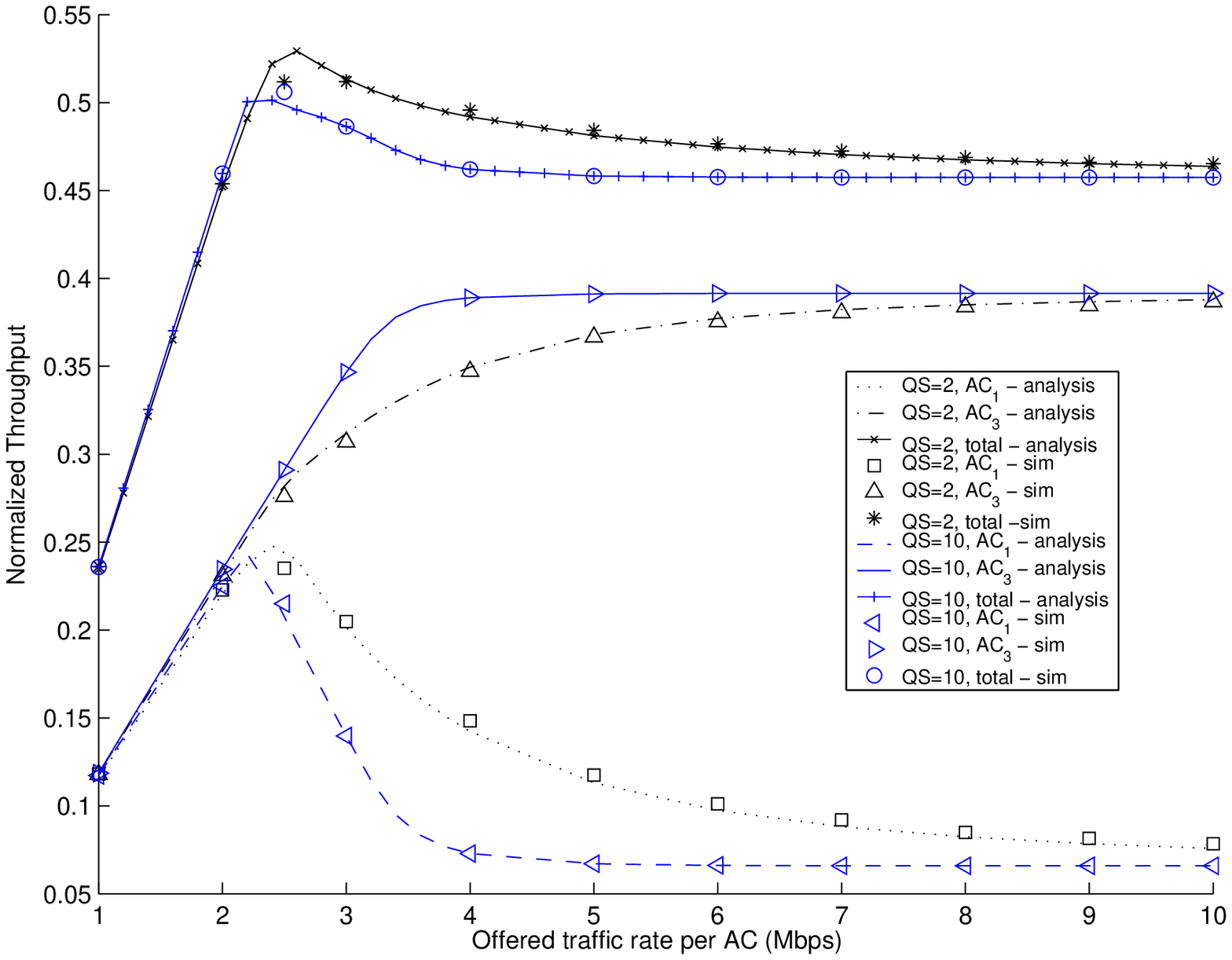} \caption{Normalized throughput
prediction of the proposed model for 2 AC heterogeneous scenario
with respect to increasing load per AC at each station and varying
MAC buffer size in basic access mode ($TXOP_{3}=0$, $TXOP_{1}=0$).
Simulation results are also added for
comparison.}\label{fig:unsat_noTXOP_varyQ}
\end{figure*}

Fig.~\ref{fig:unsat_noTXOP_varyQ} shows the differentiation of
throughput for two ACs when EDCA TXOP limits of both are set to 0
(1 packet exchange per EDCA TXOP). In this scenario, there are 5
stations for both ACs and they are transmitting to an AP. The
normalized throughput per AC as well as the total system
throughput is plotted for increasing offered load per AC. We have
carried out the analysis for maximum MAC buffer sizes of 2 packets
and 10 packets. The comparison between analytical and simulation
results shows that our model can accurately capture the linear
relationship between throughput and offered load under low loads,
the complex transition in throughput between under-loaded and
saturation regimes, and the saturation throughput. Although we do
not present here, considerable inaccuracy is observed if the
postbackoff procedure, varying collision probability among
different AIFS zones, and varying service time among different
backoff stages are not modeled correctly as proposed. The results
also present that the slot homogeneity assumption works accurately
in a non-saturated model for throughput estimation.

The proposed model can also capture the throughput variation with
respect to the size of the MAC buffer. The results reveal how
significantly the size of the MAC buffer affects the throughput in
the transition period from underloaded to highly loaded channel.
This also shows small interface buffer assumptions of previous
models
\cite{Duffy05},\cite{Shabdiz04},\cite{Shabdiz06},\cite{Tantra06}
can lead to considerable analytical inaccuracies. Although the
total throughput for the small buffer size case has higher
throughput in the transition region for the specific example, this
cannot be generalized. The reason for this is that AC$_{1}$
suffers from low throughput for $QS_{1}=10$ due to the selection
of EDCA parameters, which affects the total throughput.

It is also important to note that the throughput performance does
not differ significantly (around \%1-\%2) for buffer sizes larger
than 10 packets for the given scenarios. Therefore, we do not
include such cases in order not to complicate the figures. Since
the complexity of the mathematical solution increases with the
increasing size of the third dimension of DTMC, it may be
preferable to implement the model for smaller queue sizes when the
throughput performance is not expected to be affected by the
selection.

\begin{figure*}[p]
\centering \includegraphics[width =
1.0\linewidth]{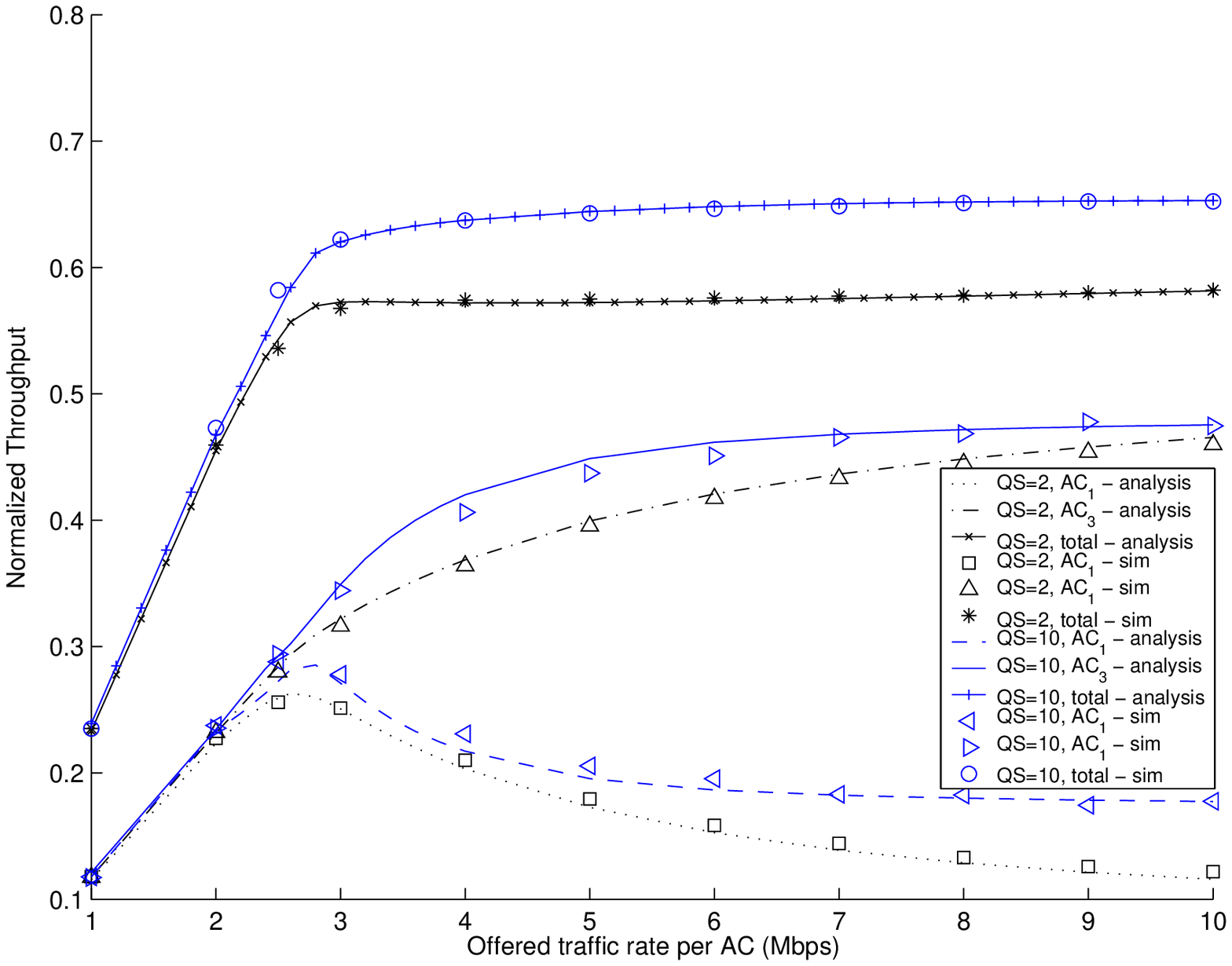} \caption{Normalized throughput
prediction of the proposed model for 2 AC heterogeneous scenario
with respect to increasing load per AC at each station and varying
MAC buffer size in basic access mode ($TXOP_{3}=1504ms$,
$TXOP_{1}=3008ms$). Simulation results are also added for
comparison.}\label{fig:unsat_TXOP_varyQ}
\end{figure*}

Fig.~\ref{fig:unsat_TXOP_varyQ} depicts the differentiation of
throughput for two ACs when EDCA TXOP limits are set to $1.504$ ms
and $3.008$ ms for high and low priority ACs respectively. For
TXOP limits, we use the suggested values for voice and video ACs
in \cite{802.11e}. It is important to note that the model works
for an arbitrary selection of the TXOP limit. According to the
selected TXOP limits, $N_{1}=5$ and $N_{2}=11$. The normalized
throughput per AC as well as the total system throughput is
plotted while increasing offered load per AC. We have done the
analysis for maximum MAC buffer sizes of 2 packets and 10 packets.
The model accurately captures the throughput for any traffic load.
As expected, increasing maximum buffer size to 10 packets
increases the throughput both in the transition and the saturation
region. Note that when more than a packet fits into EDCA TXOPs,
this decreases contention overhead which in turn increases channel
utilization and throughput (comparison of
Fig.~\ref{fig:unsat_TXOP_varyQ} with
Fig.~\ref{fig:unsat_noTXOP_varyQ}). Although corresponding results
are not presented here, the model works accurately for higher
queue sizes in the case of EDCA TXOPs as well.

\begin{figure*}[p]
\centering \includegraphics[width =
1.0\linewidth]{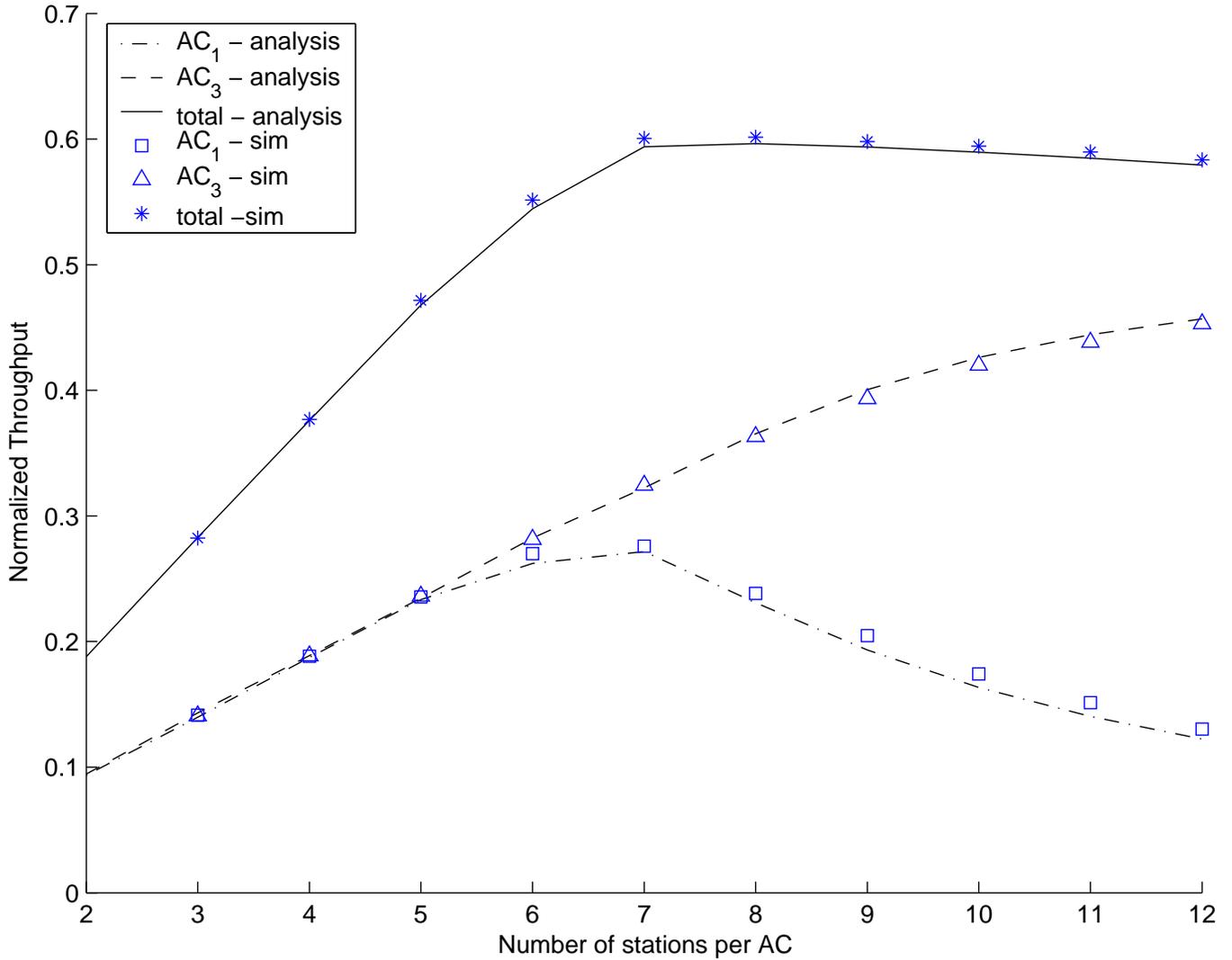} \caption{Normalized
throughput prediction of the proposed model for 2 AC heterogeneous
scenario with respect to increasing number of stations when MAC
buffer size is 10 packets and total offered load per AC is 2 Mbps
($TXOP_{3}=1504ms$, $TXOP_{1}=3008ms$). Simulation results are
also added for comparison.}\label{fig:unsat_TXOP_q10_varyn}
\end{figure*}

Fig.~\ref{fig:unsat_TXOP_q10_varyn} displays the differentiation
of throughput for two ACs when packet arrival rate is fixed to $2$
Mbps and the station number per AC is increased. We have done the
analysis for the MAC buffer size of 10 packets with EDCA TXOPs
enabled. The analytical and simulation results are well in
accordance. As the traffic load increases, the differentiation in
throughput between the ACs is observed.

\begin{figure*}[p]
\centering \includegraphics[width =
1.0\linewidth]{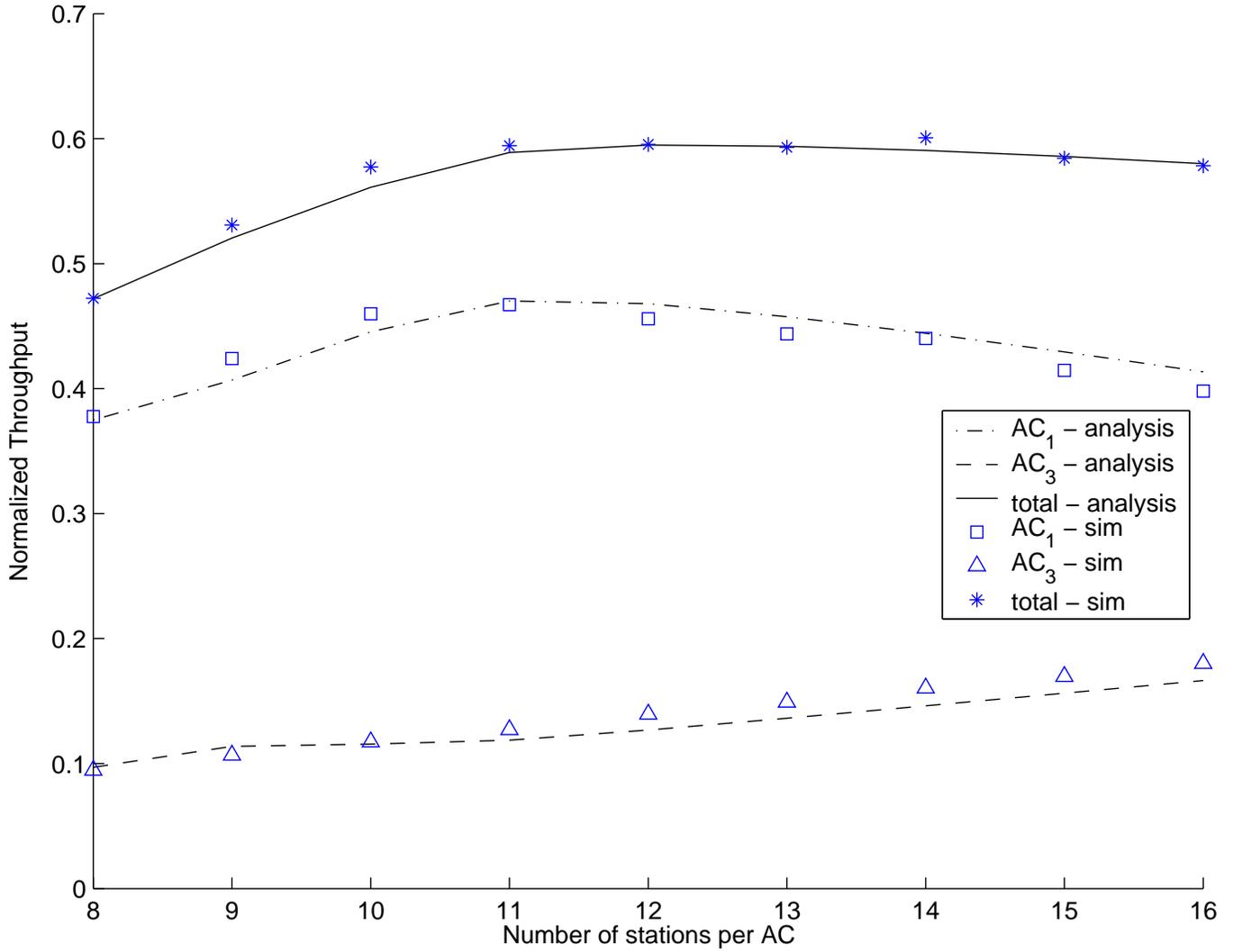} \caption{Normalized
throughput prediction of the proposed model for 2 AC heterogeneous
scenario with respect to increasing number of stations when MAC
buffer size is 10 packets ($TXOP_{3}=1504ms$, $TXOP_{1}=3008ms$).
Total offered load per AC$_{3}$ is 0.5 Mbps while total offered
load per AC$_{3}$ is 2 Mbps. Simulation results are also added for
comparison.}\label{fig:unsat_TXOP_q10_varylambdan}
\end{figure*}

Fig.~\ref{fig:unsat_TXOP_q10_varylambdan} shows the normalized
throughput for two ACs when offered load per AC is not equal. In
this scenario, we set the packet arrival rate per AC$_{1}$ to $2$
Mbps and the packet arrival rate per AC$_{3}$ to $0.5$ Mbps. The
analytical and simulation results are well in accordance. As the
traffic load increases, AC$_{3}$ maintains linear increase with
respect to offered load, while AC$_{1}$ experiences decrease in
throughput due to larger settings of AIFS and CW if the total
number of stations exceeds 22.

\begin{figure*}[p]
\centering \includegraphics[width =
1.0\linewidth]{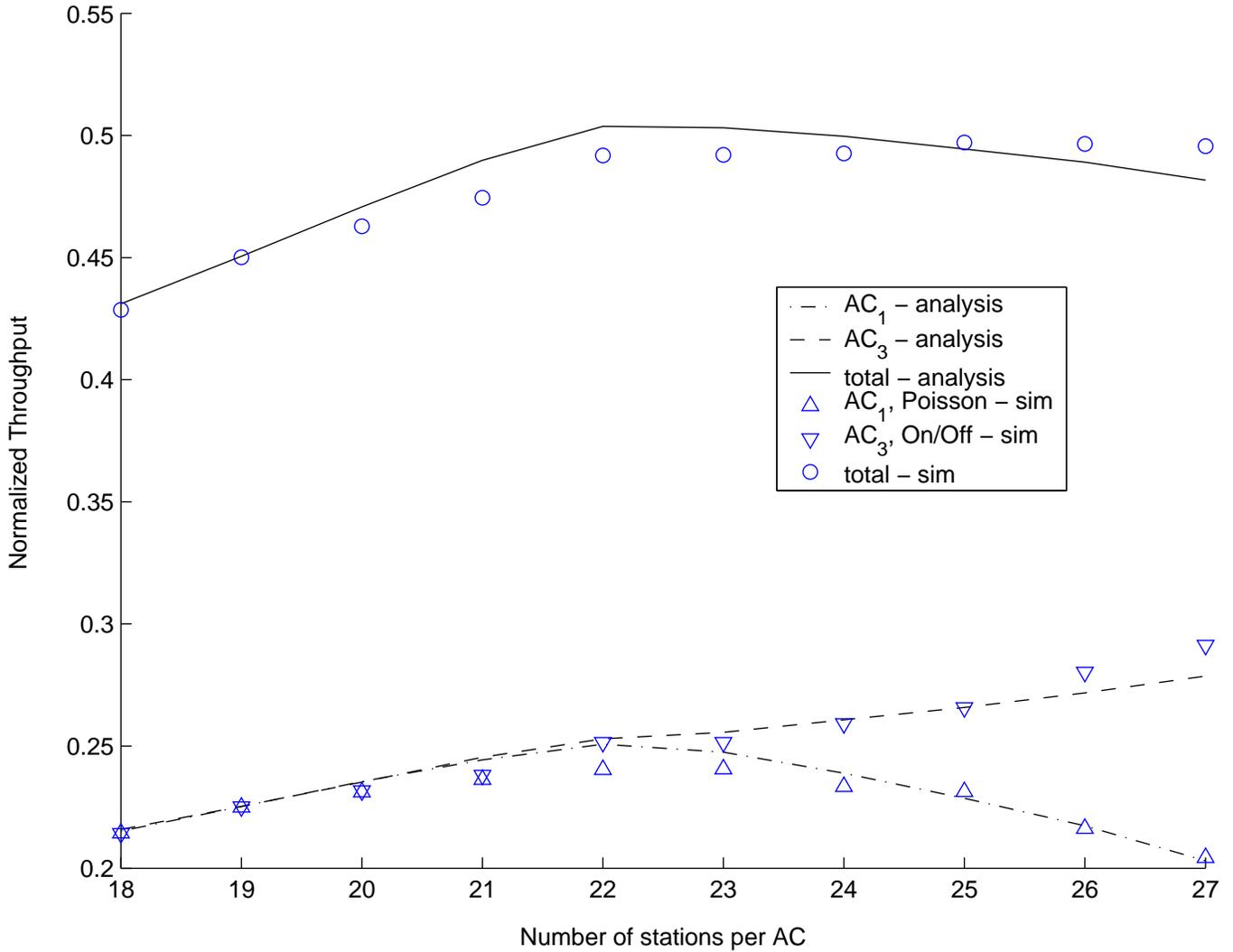} \caption{Normalized throughput
prediction of the proposed model for 2 AC heterogeneous scenario
with respect to increasing number of stations when total offered
load per AC is 0.5 Mbps ($TXOP_{3}=1504ms$, $TXOP_{1}=3008ms$).
Simulation results are also added for the scenario when AC$_{3}$
uses On/Off traffic with exponentially distributed idle and active
times both with mean 1.5s. $AC_{1}$ uses Poisson distribution for
packet arrivals. }\label{fig:unsat_varytraffic}
\end{figure*}

In the design of the model, we assume constant packet arrival
probability per state. The Poisson arrival process fits this
definition because of the independent exponentially distributed
interarrival times. We have also compared the throughput estimates
obtained from the analytical model with the simulation results
obtained using an On/Off traffic model in
Fig.~\ref{fig:unsat_varytraffic}. A similar study has first been
made for DCF in \cite{Duffy05}. We modeled the high priority with
On/Off traffic model with exponentially distributed idle and
active intervals of mean length $1.5$ s. In the active interval,
packets are generated with Constant Bit Rate (CBR). The low
priority traffic uses Poisson distributed arrivals. Note that we
leave the packet size unchanged, but normalize the packet arrival
rate according to the on/off pattern so that total offered load
remains constant to have a fair comparison. The analytical
predictions closely follow the simulation results for the given
scenario. We have observed that the predictions are more sensitive
if the transition region is entered with a few number of stations
(5 stations per AC).

\begin{figure*}[p]
\centering \includegraphics[width =
1.0\linewidth]{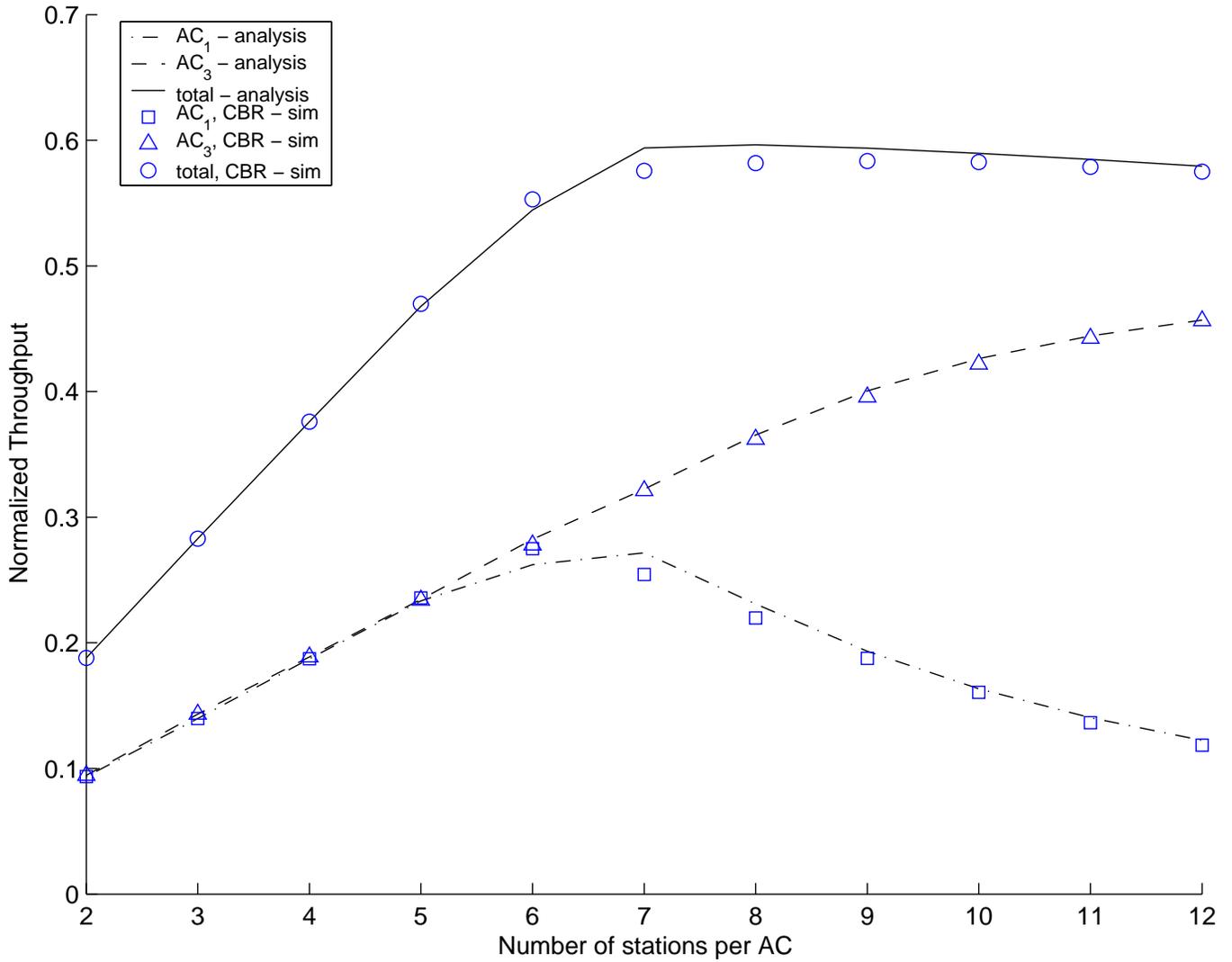} \caption{Normalized
throughput prediction of the proposed model for 2 AC heterogeneous
scenario with respect to increasing number of stations when MAC
buffer size is 10 packets and total offered load per AC is 2 Mbps
($TXOP_{3}=1504ms$, $TXOP_{1}=3008ms$). Simulation results are
also added for the scenario when both AC$_{1}$ and AC$_{3}$ uses
CBR traffic.}\label{fig:unsat_TXOP_q10_varyn_CBR}
\end{figure*}

Our model also provides a very good match in terms of the
throughput for CBR traffic. In
Fig.~\ref{fig:unsat_TXOP_q10_varyn_CBR}, we compare the throughput
prediction of the proposed model with simulations using CBR
traffic. The packet arrival rate is fixed to $2$ Mbps for both ACs
and the station number per AC is increased. MAC buffer size is 10
packets and EDCA TXOPs are enabled.

\begin{figure*}[p]
\centering \includegraphics[width =
1.0\linewidth]{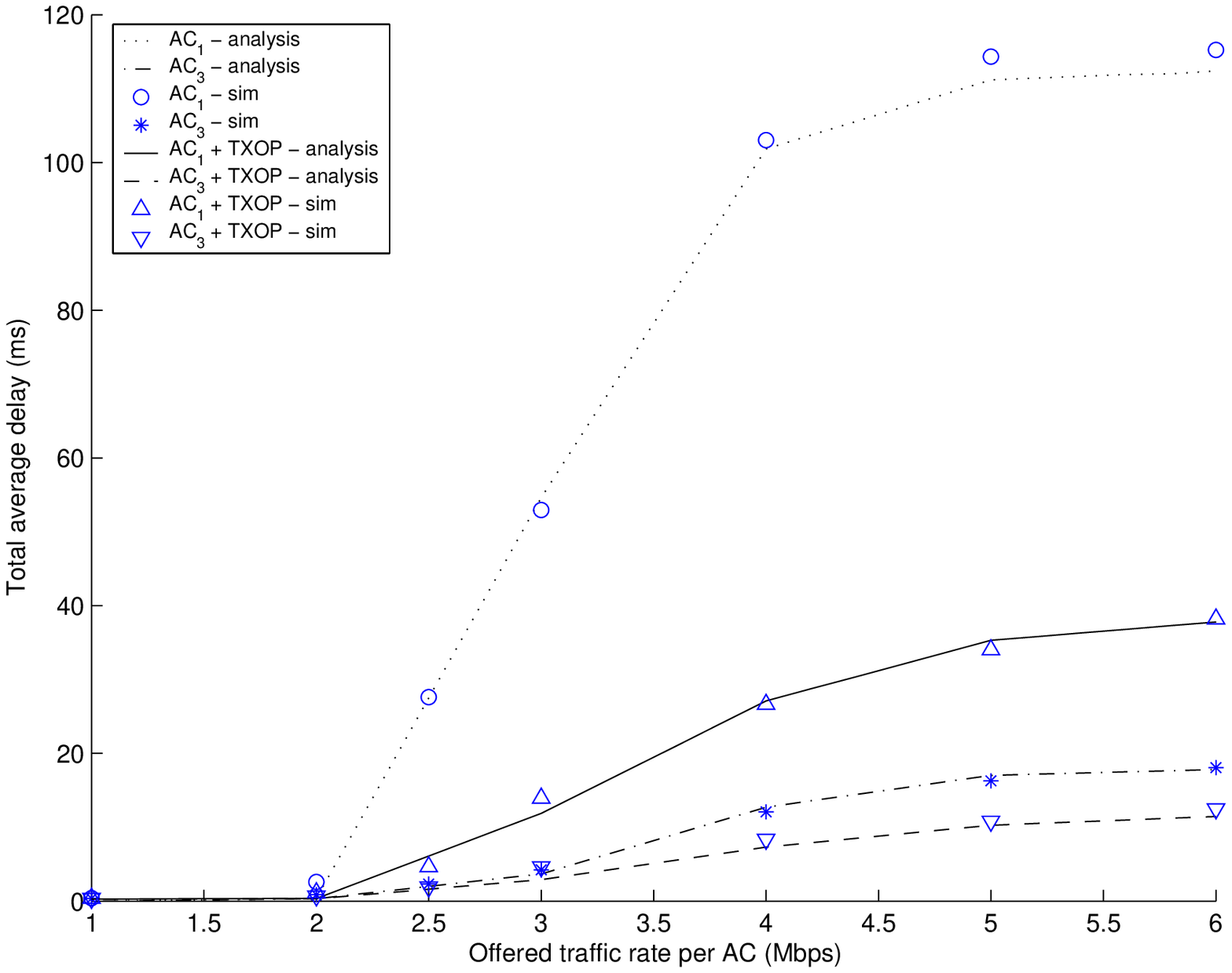} \caption{Total average delay
prediction of the proposed model for 2 AC heterogeneous scenario
with respect to increasing load per AC at each station. In the
first scenario, TXOP limits are set to 0 ms for both ACs. In the
second scenario, TXOP limits are set to $1.504$ ms and $3.008$ ms
for high and low priority ACs respectively. Simulation results are
also added for comparison.}\label{fig:unsat_delay_varyTXOP}
\end{figure*}

Fig.~\ref{fig:unsat_delay_varyTXOP} depicts the total average
packet delay with respect to increasing traffic load per AC. We
present the results for two different scenarios. In the first
scenario, TXOP limits are set to 0 ms for both ACs. In the second
scenario, TXOP limits are set to $1.504$ ms and $3.008$ ms for
high and low priority ACs respectively. The analysis is carried
out for a buffer size of 10 packets. As the results imply, the
analytical results closely follow the simulation results for both
scenarios. In the lightly loaded region, the delays are
considerably small. The increase in the transition region is
steeper when TXOP limits are 0. In the specific example, enabling
TXOPs decreases the total delay where the decrease is more
considerable for the low priority AC (due to selection of
parameters). Since the buffer size is limited, the total average
delay converges to a specific value as the load increases. Still
this limit is not of interest, since the packet loss rate at this
region is unpractically large. Note that this limit will be higher
for larger buffers. The region of interest is the start of the
transition region (between 2 Mbps and 3 Mbps for the example in
Fig.~\ref{fig:unsat_delay_varyTXOP}). On the other hand, we also
display other data points to show the performance of the model for
the whole load span.

\begin{figure*}[p]
\centering \includegraphics[width =
1.0\linewidth]{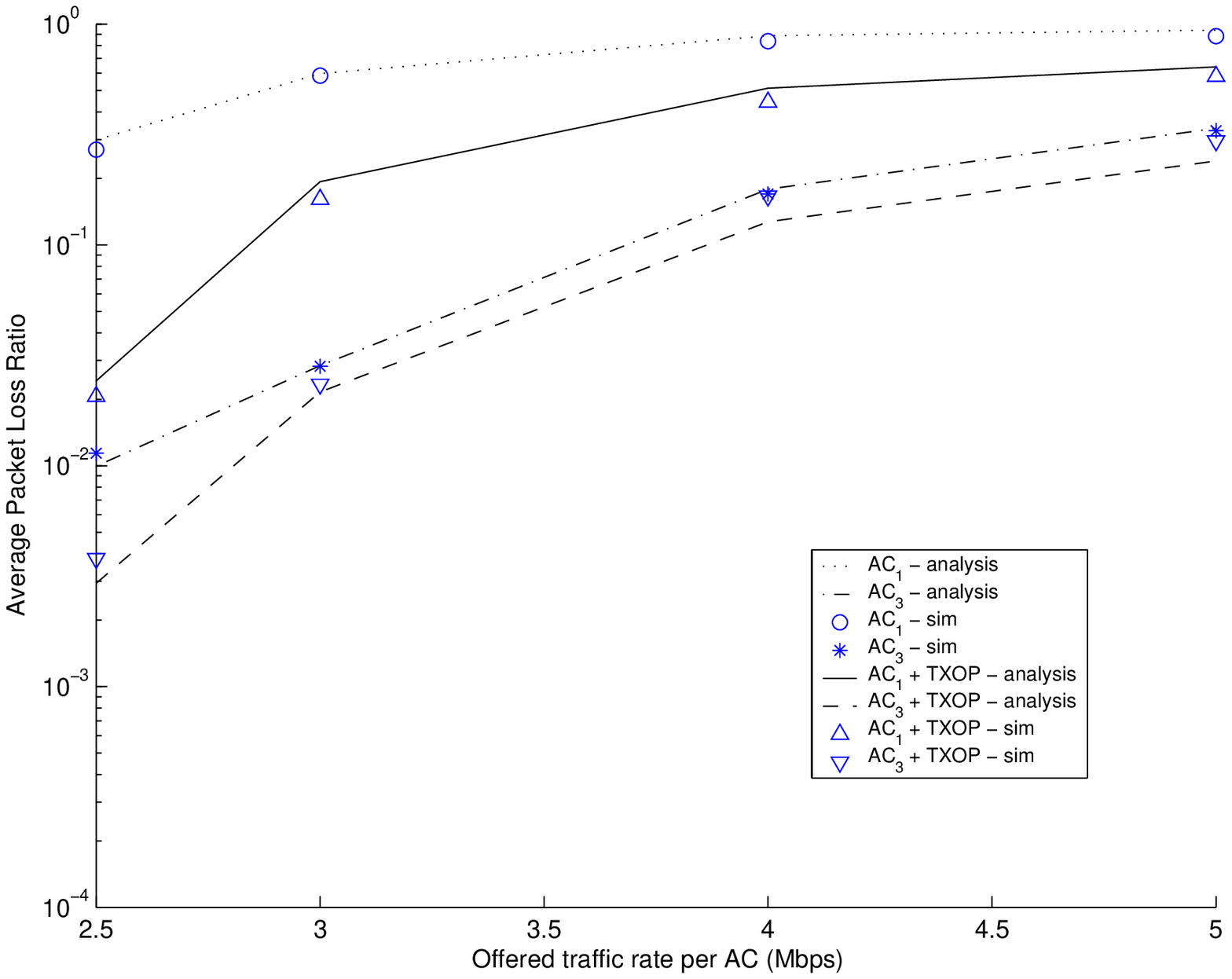} \caption{Average packet
loss ratio prediction of the proposed model for 2 AC heterogeneous
scenario with respect to increasing load per AC at each station.
In the first scenario, TXOP limits are set to 0 ms for both ACs.
In the second scenario, TXOP limits are set to $1.504$ ms and
$3.008$ ms for high and low priority ACs respectively. Simulation
results are also added for
comparison.}\label{fig:unsat_q10_varyload_plr}
\end{figure*}

Fig.~\ref{fig:unsat_q10_varyload_plr} depicts the average packet
loss ratio with respect to increasing traffic load per AC. We
present the results for two different scenarios. In the first
scenario, TXOP limits are set to 0 ms for both ACs. In the second
scenario, TXOP limits are set to $1.504$ ms and $3.008$ ms for
high and low priority ACs respectively. The analysis is carried
out for a buffer size of 10 packets. As the results imply, the
analytical results closely follow the simulation results for both
scenarios. Although it is not presented in
Fig.~\ref{fig:unsat_q10_varyload_plr}, the packet loss ratio drops
exponentially to 0 when the offered load per AC is lower than 2.5
Mbps.

The results presented in this paper fixes the AIFS and CW
parameters for each AC. The results are compared for different
TXOP values at varying traffic load. Therefore, the presented
results can mainly indicate the effects of TXOP on the maximum
throughput. The model can also be used in order to investigate the
effects of AIFS and CW on the maximum throughput.

As the comparison of Fig.~\ref{fig:unsat_noTXOP_varyQ} and
Fig.~\ref{fig:unsat_TXOP_varyQ} reveals, the total throughput can
be maximized with the introduction of EDCA TXOPs which enable
multiple frame transmissions in one channel access (note that MAC
buffer sizes for each AC should be equal to or larger than the
number of packets that can fit to the AC-specific TXOP in order to
efficiently utilize each TXOP gained). EDCA TXOPs decrease the
channel contention overhead and the ACs can efficiently utilize
the resources. Note also that the effects of EDCA TXOPs in the
lightly loaded region is marginal compared to highly loaded
region. This is expected since the MAC queues do not build up in
the lightly loaded scenario where stations usually have just one
packet to send at their access to the channel.

As Fig.~\ref{fig:unsat_noTXOP_varyQ} shows the saturation
throughput is usually less than the maximum throughput that can be
obtained. This is also observed for DCF in \cite{Bianchi00}.
Similarly, in
Fig.~\ref{fig:unsat_TXOP_q10_varyn}-Fig.~\ref{fig:unsat_TXOP_q10_varyn_CBR},
the total throughput slightly decreases as the total load
increases. As the load in the system increases the collision
overhead becomes significant which decreases the total channel
utilization. On the other hand, as also discussed in
\cite{Bianchi00}, the point where the maximum throughput is
observed is unstable in a random access system. Therefore, a good
admission control algorithm should be defined to operate the
system at the point right before the lightly loaded to highly
loaded transition region starts.

\section{Conclusion}\label{sec:conclusion}

We have presented an accurate Markov model for analytically
calculating the EDCA throughput and delay for the whole traffic
load range from a lightly loaded non-saturated channel to a
heavily congested saturated medium. The presented model shows the
accuracy of the homogeneous slot assumption (constant collision
and transmission probability at an arbitrary backoff slot) that is
extensively studied in saturation scenarios for the whole traffic
range. The presented model accurately captures the linear
relationship between throughput and offered load under low loads
and the limiting behavior of throughput at saturation.

The key contribution of this paper is that the model accounts for
all of the differentiation mechanisms EDCA proposes. The
analytical model can incorporate any selection of AC-specific
AIFS, CW, and TXOP values for any number of ACs. The model also
considers varying collision probabilities at different contention
zones which provides accurate AIFS differentiation analysis.
Although not presented explicitly in this paper, it is
straightforward to extend the presented model for scenarios where
the stations run multiple ACs (virtual collisions may take place)
or RTS/CTS protection mechanism is used. The approximations made
for the sake of DTMC simplicity and symmetry may also be removed
easily for increased accuracy, although they are shown to be
highly accurate.

We also show that the MAC buffer size affects the EDCA performance
significantly between underloaded and saturation regimes
(including saturation) especially when EDCA TXOPs are enabled. The
presented model captures this complex transition accurately. This
analysis also points out the fact that including an accurate queue
treatment is vital. Incorporating MAC queue states also enables
EDCA TXOP analysis so that the EDCA TXOP continuation process is
modeled in considerable detail. To the authors' knowledge this is
the first demonstration of an analytic model including EDCA TXOP
procedure for finite load.

It is also worth noting that our model can easily be simplified to
model DCF behavior. Moreover, after modifying our model
accordingly, the throughput analysis for the infrastructure WLAN
where there are transmissions both in the uplink and downlink can
be performed (note that in a WLAN downlink traffic load may
significantly differ from uplink traffic load).

Although the Markov analysis assumes the packets are generated
according to Poisson process, the comparison with simulation
results shows that the throughput analysis is valid for a range of
traffic types such as CBR and On/Off traffic (On/Off traffic model
is a widely used model for voice and telnet traffic).

The non-existence of a closed-form solution for the Markov model
limits its practical use. On the other hand, the accurate
saturation throughput analysis can highlight the strengths and the
shortcomings of EDCA for varying scenarios and can provide
invaluable insights. The model can effectively assist EDCA
parameter adaptation or a call admission control algorithm for
improved QoS support in the WLAN.


\bibliographystyle{IEEEtran}
\bibliography{IEEEabrv,C:/INANCINAN/bibliography/standards,C:/INANCINAN/bibliography/HCCA,C:/INANCINAN/bibliography/simulations,C:/INANCINAN/bibliography/channel,C:/INANCINAN/bibliography/books,C:/INANCINAN/bibliography/EDCAanalysis,C:/INANCINAN/bibliography/mypapers,C:/INANCINAN/bibliography/myreports}

\begin{thebibliography}{10}
\providecommand{\url}[1]{#1}
\csname url@rmstyle\endcsname
\providecommand{\newblock}{\relax}
\providecommand{\bibinfo}[2]{#2}
\providecommand\BIBentrySTDinterwordspacing{\spaceskip=0pt\relax}
\providecommand\BIBentryALTinterwordstretchfactor{4}
\providecommand\BIBentryALTinterwordspacing{\spaceskip=\fontdimen2\font plus
\BIBentryALTinterwordstretchfactor\fontdimen3\font minus
  \fontdimen4\font\relax}
\providecommand\BIBforeignlanguage[2]{{%
\expandafter\ifx\csname l@#1\endcsname\relax
\typeout{** WARNING: IEEEtran.bst: No hyphenation pattern has been}%
\typeout{** loaded for the language `#1'. Using the pattern for}%
\typeout{** the default language instead.}%
\else
\language=\csname l@#1\endcsname
\fi
#2}}

\bibitem{802.11}
\emph{{IEEE Standard 802.11: Wireless {LAN} medium access control (MAC) and
  physical layer (PHY) specifications}}, {IEEE 802.11} Std., 1999.

\bibitem{802.11e}
\emph{{IEEE Standard 802.11: Wireless {LAN} medium access control (MAC) and
  physical layer (PHY) specifications: Medium access control (MAC) Quality of
  Service (QoS) Enhancements}}, {IEEE 802.11e} Std., 2005.

\bibitem{Bianchi00}
G.~Bianchi, ``{Performance Analysis of the IEEE 802.11 Distributed Coordination
  Function},'' \emph{{IEEE} Trans. Commun.}, pp. 535--547, March 2000.

\bibitem{Cali98}
F.~Cali, M.~Conti, and E.~Gregori, ``{IEEE 802.11 Wireless LAN: Capacity
  Analysis and Protocol Enhancement},'' in \emph{Proc. IEEE Infocom '98}, March
  1998.

\bibitem{Cali00}
------, ``{Dynamic Tuning of the IEEE 802.11 Protocol to Achieve a Theoretical
  Throughput Limit},'' \emph{IEEE/ACM Trans. Netw.}, pp. 785--799, December
  2000.

\bibitem{Tay01}
J.~C. Tay and K.~C. Chua, ``{A Capacity Analysis for the IEEE 802.11 MAC
  Protocol},'' \emph{Wireless Netw.}, pp. 159--171, July 2001.

\bibitem{Medepalli05}
K.~Medepalli and F.~A. Tobagi, ``{Throughput Analysis of IEEE 802.11 Wireless
  LANs using an Average Cycle Time Approach},'' in \emph{Proc. IEEE Globecom
  '05}, November 2005.

\bibitem{Hui06}
J.~Hui and M.~Devetsikiotis, ``{Metamodeling of Wi-Fi Performance},'' in
  \emph{Proc. IEEE ICC '06}, June 2006.

\bibitem{Xiao04}
Y.~Xiao, ``{An Analysis for Differentiated Services in IEEE 802.11 and IEEE
  802.11e Wireless LANs},'' in \emph{Proc. IEEE ICDCS '04}, March 2004.

\bibitem{Xiao05}
------, ``{Performance Analysis of Priority Schemes for IEEE 802.11 and IEEE
  802.11e Wireless LANs},'' \emph{{IEEE} Trans. Wireless Commun.}, pp.
  1506--1515, July 2005.

\bibitem{Kong04}
Z.~Kong, D.~H.~K. Tsang, B.~Bensaou, and D.~Gao, ``{Performance Analysis of the
  IEEE 802.11e Contention-Based Channel Access},'' \emph{{IEEE} J. Select.
  Areas Commun.}, pp. 2095--2106, December 2004.

\bibitem{Robinson04}
J.~W. Robinson and T.~S. Randhawa, ``{Saturation Throughput Analysis of IEEE
  802.11e Enhanced Distributed Coordination Function},'' \emph{{IEEE} J.
  Select. Areas Commun.}, pp. 917--928, June 2004.

\bibitem{Robinson04_2}
------, ``{A Practical Model for Transmission Delay of IEEE 802.11e Enhanced
  Distributed Channel Access},'' in \emph{Proc. IEEE PIMRC '04}, September
  2004.

\bibitem{Hui04}
J.~Hui and M.~Devetsikiotis, ``{Performance Analysis of IEEE 802.11e EDCA by a
  Unified Model},'' in \emph{Proc. IEEE Globecom '04}, December 2004.

\bibitem{Hui05}
------, ``{A Unified Model for the Performance Analysis of IEEE 802.11e
  EDCA},'' \emph{{IEEE} Trans. Commun.}, pp. 1498--1510, September 2005.

\bibitem{Zhu05}
H.~Zhu and I.~Chlamtac, ``{Performance Analysis for IEEE 802.11e EDCF Service
  Differentiation},'' \emph{{IEEE} Trans. Wireless Commun.}, pp. 1779--1788,
  July 2005.

\bibitem{Inan07_ICC}
I.~Inan, F.~Keceli, and E.~Ayanoglu, ``{Saturation Throughput Analysis of the
  802.11e Enhanced Distributed Channel Access Function},'' to appear in Proc.
  IEEE ICC '07.

\bibitem{Tao04}
Z.~Tao and S.~Panwar, ``{An Analytical Model for the IEEE 802.11e Enhanced
  Distributed Coordination Function},'' in \emph{Proc. IEEE ICC '04}, May 2004.

\bibitem{Tao06}
------, ``{Throughput and Delay Analysis for the IEEE 802.11e Enhanced
  Distributed Channel Access},'' \emph{{IEEE} Trans. Commun.}, pp. 596--602,
  April 2006.

\bibitem{Zhao02}
J.~Zhao, Z.~Guo, Q.~Zhang, and W.~Zhu, ``{Performance Study of MAC for Service
  Differentiation in IEEE 802.11},'' in \emph{Proc. IEEE Globecom '02},
  November 2002.

\bibitem{Banchs05}
A.~Banchs and L.~Vollero, ``{A Delay Model for IEEE 802.11e EDCA},''
  \emph{{IEEE} Commun. Lett.}, pp. 508--510, June 2005.

\bibitem{Banchs06}
------, ``{Throughput Analysis and Optimal Configuration of IEEE 802.11e
  EDCA},'' \emph{Comp. Netw.}, pp. 1749--1768, August 2006.

\bibitem{Chen03}
Y.~Chen, Q.-A. Zeng, and D.~P. Agrawal, ``{Performance Analysis of IEEE 802.11e
  Enhanced Distributed Coordination Function},'' in \emph{Proc. IEEE ICON '03},
  September 2003.

\bibitem{Kuo03}
Y.-L. Kuo, C.-H. Lu, E.~H.-K. Wu, G.-H. Chen, and Y.-H. Tseng, ``{Performance
  Analysis of the Enhanced Distributed Coordination Function in the IEEE
  802.11e},'' in \emph{Proc. IEEE VTC '03 - Fall}, October 2003.

\bibitem{Lin06}
Y.~Lin and V.~W. Wong, ``{Saturation Throughput of IEEE 802.11e EDCA Based on
  Mean Value Analysis},'' in \emph{Proc. IEEE WCNC '06}, April 2006.

\bibitem{Kleinrock75}
L.~Kleinrock, \emph{{Queueing Systems}}.\hskip 1em plus 0.5em minus 0.4em\relax
  John Wiley and Sons, 1975.

\bibitem{Duffy05}
K.~Duffy, D.~Malone, and D.~J. Leith, ``{Modeling the 802.11 Distributed
  Coordination Function in Non-Saturated Conditions},'' \emph{{IEEE} Commun.
  Lett.}, pp. 715--717, August 2005.

\bibitem{Shabdiz04}
F.~Alizadeh-Shabdiz and S.~Subramaniam, ``{Analytical Models for Single-Hop and
  Multi-Hop Ad Hoc Networks},'' in \emph{Proc. ACM Broadnets '04}, October
  2004.

\bibitem{Shabdiz06}
------, ``{Analytical Models for Single-Hop and Multi-Hop Ad Hoc Networks},''
  \emph{Mobile Networks and Applications}, pp. 75--90, February 2006.

\bibitem{Cantieni05}
G.~R. Cantieni, Q.~Ni, C.~Barakat, and T.~Turletti, ``{Performance Analysis
  under Finite Load and Improvements for Multirate 802.11},'' \emph{Comp.
  Commun.}, pp. 1095--1109, June 2005.

\bibitem{Li05}
B.~Li and R.~Battiti, ``{Analysis of the IEEE 802.11 DCF with Service
  Differentiation Support in Non-Saturation Conditions},'' in \emph{QoFIS '04},
  September 2004.

\bibitem{Engelstad06}
P.~E. Engelstad and O.~N. Osterbo, ``{Analysis of the Total Delay of IEEE
  802.11e EDCA and 802.11 DCF},'' in \emph{Proc. IEEE ICC '06}, June 2006.

\bibitem{Zaki04}
A.~N. Zaki and M.~T. El-Hadidi, ``{Throughput Analysis of IEEE 802.11 DCF Under
  Finite Load Traffic},'' in \emph{Proc. First International Symposium on
  Control, Communications and Signal Processing}, 2004.

\bibitem{Tickoo04_1}
O.~Tickoo and B.~Sikdar, ``{Queueing Analysis and Delay Mitigation in IEEE
  802.11 Random Access MAC based Wireless Networks},'' in \emph{Proc. IEEE
  Infocom '04}, March 2004.

\bibitem{Tickoo04_2}
------, ``{A Queueing Model for Finite Load IEEE 802.11 Random Access MAC},''
  in \emph{Proc. IEEE ICC '04}, June 2004.

\bibitem{XChen06}
X.~Chen, H.~Zhai, X.~Tian, and Y.~Fang, ``{Supporting QoS in IEEE 802.11e
  Wireless LANs},'' \emph{{IEEE} Trans. Wireless Commun.}, pp. 2217--2227,
  August 2006.

\bibitem{Lee06}
W.~Lee, C.~Wang, and K.~Sohraby, ``{On Use of Traditional M/G/1 Model for IEEE
  802.11 DCF in Unsaturated Traffic Conditions},'' in \emph{Proc. IEEE WCNC
  '06}, May 2006.

\bibitem{Medepalli05_2}
K.~Medepalli and F.~A. Tobagi, ``{System Centric and User Centric Queueing
  Models for IEEE 802.11 based Wireless LANs},'' in \emph{Proc. IEEE Broadnets
  '05}, October 2005.

\bibitem{Foh02}
C.~H. Foh and M.~Zukerman, ``{A New Technique for Performance Evaluation of
  Random Access Protocols},'' in \emph{Proc. European Wireless '02}, February
  2002.

\bibitem{Tantra06}
J.~W. Tantra, C.~H. Foh, I.~Tinnirello, and G.~Bianchi, ``{Analysis of the IEEE
  802.11e EDCA Under Statistical Traffic},'' in \emph{Proc. IEEE ICC '06}, June
  2006.

\bibitem{Mangold02}
S.~Mangold, S.~Choi, P.~May, and G.~Hiertz, ``{IEEE 802.11e - Fair Resource
  Sharing Between Overlapping Basic Service Sets},'' in \emph{Proc. IEEE PIMRC
  '02}, September 2002.

\bibitem{Suzuki06}
T.~Suzuki, A.~Noguchi, and S.~Tasaka, ``{Effect of TXOP-Bursting and
  Transmission Error on Application-Level and User-Level QoS in Audio-Video
  Transmission with 802.11e EDCA},'' in \emph{Proc. IEEE PIMRC '06}, September
  2006.

\bibitem{Tinnirello05_2}
I.~Tinnirello and S.~Choi, ``{Efficiency Analysis of Burst Transmissions with
  Block ACK in Contention-Based 802.11e WLANs},'' in \emph{Proc. IEEE ICC '05},
  May 2005.

\bibitem{Tinnirello05}
------, ``{Temporal Fairness Provisioning in Multi-Rate Contention-Based
  802.11e WLANs},'' in \emph{Proc. IEEE WoWMoM '05}, June 2005.

\bibitem{Peng06}
F.~Peng, H.~M. Alnuweiri, and V.~C.~M. Leung, ``{Analysis of Burst Transmission
  in IEEE 802.11e Wireless LANs},'' in \emph{Proc. IEEE ICC '06}, June 2006.

\bibitem{ns2}
\BIBentryALTinterwordspacing
(2006) {The Network Simulator, ns-2}. [Online]. Available:
  \url{http://www.isi.edu/nsnam/ns}
\BIBentrySTDinterwordspacing

\bibitem{ourcode}
\BIBentryALTinterwordspacing
{IEEE 802.11e HCF MAC model for ns-2.28}. [Online]. Available:
  \url{http://newport.eecs.uci.edu/$\sim$fkeceli/ns.htm}
\BIBentrySTDinterwordspacing

\bibitem{802.11g}
\emph{{IEEE Standard 802.11: Wireless {LAN} medium access control (MAC) and
  physical layer (PHY) specifications: Further Higher Data Rate Extension in
  the 2.4 GHz Band}}, {IEEE 802.11g} Std., 2003.

\end{thebibliography}



\clearpage

\clearpage

\clearpage

\clearpage

\clearpage

\clearpage

\clearpage

\clearpage

\clearpage

\clearpage

\clearpage


\end{document}